%% file: main.tex
\documentclass[a4paper,11pt]{article}
\pdfoutput=1 
\usepackage{jcappub}
\usepackage[T1]{fontenc} 

\usepackage[dvipsnames]{xcolor}

\usepackage{booktabs}

\usepackage{mathtools}
\usepackage{physics}
\usepackage{xspace}

\newcommand{\github}{{\tt GitHub}\xspace}
\newcommand{\baofitwtheta}{{\tt BAOfit\_wtheta}\xspace}

\newcommand{\abacus}{{\tt Abacus}\xspace}
\newcommand{\abacussummit}{{\sc AbacusSummit}\xspace}
\newcommand{\altmtl}{{\tt altMTL}\xspace}

\newcommand{\wvec}{{\bf w}}  
\newcommand{\Cmat}{\mathsf{C}}  
\newcommand{\Ccorr}{\hat{\mathsf{C}}}

\title{Angular Baryon Acoustic Oscillations with DESI DR1}

\author[a,b,1]{A.~Paopiamsap,\note{Corresponding author.}}
\author[c]{J.~Mena-Fernández,}
\author[a,b,d]{H.~Gil-Marín}
\author[a,e]{and L.~Verde}

\affiliation[a]{Institut de Ciències del Cosmos (ICCUB), Universitat de Barcelona (UB), c. Martí i Fran-
quès, 1, 08028 Barcelona, Spain}
\affiliation[b]{Departament de Física Quàntica i Astrofísica, Universitat de Barcelona, Martí i Franquès 1, E08028 Barcelona, Spain}
\affiliation[c]{Aix Marseille Université, CNRS/IN2P3, CPPM, Marseille, France}
\affiliation[d]{Institut d’Estudis Espacials de Catalunya (IEEC), c/ Esteve Terradas 1, Edifici RDIT, Campus PMT-UPC, 08860 Castelldefels, Spain}
\affiliation[e]{Institució Catalana de Recerca i Estudis Avançats, Passeig de Lluís Companys, 23, 08010
Barcelona, Spain}

\emailAdd{anyap@icc.ub.edu,menafernandez@cppm.in2p3.fr, hectorgil@icc.ub.edu, liciaverde@icc.ub.edu}

\abstract{\input{abstract}}

\begin{document}
\maketitle
\flushbottom

\section{Introduction}
\label{sec:intro}

Baryon Acoustic Oscillations (BAO) constitute an observational pillar in modern cosmology, providing a robust standard ruler to constrain the expansion history of the Universe and test cosmological models \cite{Eisenstein:1997ik,Blake:2003rh, Seo:2003pu}. The feature originates from acoustic waves in the tightly coupled photon–baryon plasma of the early Universe. At the drag epoch, the baryonic counterpart remained stalled, imprinting the sound horizon as a characteristic scale in the late-time matter distribution. This scale is determined by well-understood early-Universe physics and calibrated with Cosmic Microwave Background (CMB) observations, making it a notably clean and low-systematics probe for late-time cosmology \cite{2009PhRvD..80f3508P,Eisenstein:2006nj,Planck:2018vyg}. 

Decades of sustained efforts in expanding survey volume, controlling observational systematics and refining theoretical modelling have allowed the BAO to evolve from a marginal statistical detection into a precision cosmological probe. The first detections in 2005 of the BAO feature in the galaxy distribution were achieved by the Sloan Digital Sky Survey (SDSS) Luminous Red Galaxies (LRGs) at $z \sim 0.35$ \cite{eisenstein2005detection} and the Two-Degree Field Galaxy Redshift Survey (2dFGRS) at $z \sim 0.2$ \cite{Cole2005The2G}. The precision achieved was at the $\sim 3\%$ level, establishing the foundation for BAO analyses in larger subsequent surveys. These surveys extended these measurements across cosmic time: Six-Degree Field Galaxy Redshift Survey (6dFGS \cite{beutler20116df}) anchored the distance scale at low redshift $(z \sim 0.1)$, WiggleZ \cite{WiggleZ:2012sek} increased the range up to $z < 1$, and the Baryon Oscillation Spectroscopic Survey (BOSS, \cite{Alam2017boss}) and its successor extended BOSS (eBOSS, \cite{Alam2021eboss}) delivered percent-level constraints while broadening the tracer set to include the clustering of LRGs, emission line galaxies (ELGs) and quasar samples (QSOs), allowing for a broad redshift-range BAO signal ($0.2<z<2.2$). Additionally, the Lyman-$\alpha$ forest signal was exploited to obtain measurements at even higher redshift ranges ($2<z<4$).

The current state-of-the-art is the Dark Energy Spectroscopic Instrument (DESI), a stage IV Dark Energy experiment that has been designed to extract BAO measurements with sub-percent precision as one of its primary science targets \cite{DESI:2024DR1, DESI:2025DR2}. DESI is an ongoing eight-year survey that covers 17,000 square degrees of the sky over a broad redshift range of $0 < z < 3.5$. The first DESI data release (DR1) in 2025 \cite{DESI:2025fxa} includes 13 months of observations from May 2021 up until June 2022, and marks the largest spectroscopic redshift catalogue released to date. The DESI DR1 contains $\sim 18.7$ million unique objects reported with high confidence, of which $ \sim 13.1$ million are galaxies and $\sim 1.6$ million are quasars. Within its first year of observations alone, DESI DR1 assembled more high-confidence extragalactic redshifts than all previous 3D spectroscopic surveys combined, surveying a cosmological volume many times larger than its predecessors, with an LRG number density roughly 25$\%$ above BOSS \cite{DESI:2024aax}. Over its lifetime, DESI aims to collect a total of $\sim 63.9$ million galaxy and quasar redshifts. 

The statistical power of DESI is conventionally exploited through 3D clustering measurements using the galaxy power spectrum or the two-point correlation function in comoving coordinates \cite{beutler20116df, WiggleZ:2012sek,Alam2017boss,Alam2021eboss, DESI:2024DR1, DESI:2025DR2}. The comoving coordinates used in a 3D clustering analysis are themselves a model-dependent mapping from the direct observables, angular positions and redshifts, by converting redshifts into distances through a fiducial cosmology, also known as ``catalogue'' cosmology. This dependence has been studied extensively and shown to be a controlled systematic at current precision \cite{Sherwin:2018wbu, Carter:2019ulk, Sanz-Wuhl:2024uvi, DESI:2024ude, DESI:2025gsp}. In particular, the motivation to study the BAO using the two-point angular correlation function (2PACF) in this work is due to its unique complementary position. Firstly, measuring the clustering as a function of angular separation within narrow redshift shells does not require a catalogue cosmology to convert observed redshifts into line-of-sight distances. Because the angular (also known as 2D) and 3D analyses can be derived from the same survey but differ precisely at the coordinate-transformation stage, comparing both results constitutes an indirect null test of the catalogue cosmology underlying the 3D pipeline. The relevance of such a test has grown with recent reports of mild tensions between the 2D and 3D BAO results \cite{Favale:2024sdq, Ruchika:2024lgi, Bernui:2023byc, Pantos:2026rpe, Dwivedi:2024okk}. Secondly, using the 2PACF has an increasingly practical motivation. Photometric surveys lack the redshift precision to resolve clustering along the line-of-sight and therefore have no access to the 3D BAO. In this case, tomographic angular clustering is no longer an alternative technique but the only avenue to access the standard ruler. An example of an existing analysis has been done using Dark Energy Survey (DES) data \cite{DES:20242dbao,mena2026dark} and Dark Energy Camera Legacy Survey (DECaLS) data \cite{Song:2024taz}. As the Vera C. Rubin Legacy Survey of Space and Time (LSST) \cite{lsst,LSSTDarkEnergyScience:2018jkl} and Euclid \cite{Euclid:2024yrr} reach maturity, a spectroscopic angular BAO measurement serves at once as a stand-alone cosmological probe and as a complementary analysis for photometric surveys against which the angular approach can be validated.

The methodology for extracting the angular BAO has been extensively used and studied in the past, both with photometric surveys \cite{Sanchez:2010zg, Song:2024taz, DES:20242dbao, sdssdr8photo,mena2026dark} and with spectroscopic surveys \cite{Carvalho:2015ica, Albornoz:2014mra, Albornoz:2016lsx, Alcaniz:2016ryy, deCarvalho:2017xye, Carvalho:2017tuu, deCarvalho:2021azj, Menote:2021jaq, Avila:2025qxc, Ferreira:2025pea}. Realising this angular clustering tomography requires thin redshift shells, which, as many of the previous works stated, have been a common obstacle due to shot-noise limitations or photometric redshift errors. DESI DR1 has made this present analysis particularly timely. Both the redshift precision and the sample density are jointly sufficient to divide the tracer samples into thin tomographic redshift shells while retaining the statistics required for a meaningful angular measurement in each. This opens the possibility of an angular BAO analysis at a resolution that previous surveys could not support. To extract the angular BAO scale, we implement a template-based fitting method as we will explain in Section \ref{sec:methods}, as previously adopted in photometric analyses \cite{sdssdr8photo, DES:20242dbao, mena2026dark}, and follow a pipeline similar to that of \cite{mena2026dark}. This differs from the empirical fitting approach used in most spectroscopic angular BAO analyses, including \cite{Carvalho:2015ica, Carvalho:2017tuu, deCarvalho:2021azj} and the most recent analysis of the DESI Bright Galaxy Survey \cite{Ferreira:2025pea}. However, utilising the full precision of the data depends on the careful control of a number of systematics, which we address in the remainder of this work.

This paper is organised as follows. In Section \ref{sec:data}, we detail the data and mock catalogues and specifically we highlight the tracer samples used for our official results. In Section \ref{sec:methods}, we describe the methodologies used in this work. In particular, we describe the choice of redshift and angular configurations used for the measurements of the two-point angular correlation function, the construction of the BAO template and its treatment of the linear galaxy bias, and the fitting procedure used to extract the angular BAO scale. Section \ref{sec:cov} presents the estimation of the covariance matrix. In Section \ref{sec:res}, we present our results: the angular correlation function measurements and best-fit models, the significance of the BAO detection, and how our measurements compare with external datasets. In Section \ref{sec:cosmoinf}, we use the angular BAO measurements to provide constraints on $\Lambda {\rm CDM}$ parameters. Finally, we summarise our conclusions in Section \ref{sec:conc}. We refer the reader to Appendix \ref{app:pipeval} for the pipeline validation on mocks, Appendix \ref{app:individualfit} for the BAO fits on individual redshift bins and Appendix \ref{app:offblock} for the tests done for the quasar covariance detailed in Section \ref{sec:cov}.

\section{Data and mock catalogues} \label{sec:data}

We use the publicly available DESI DR1 large-scale structure catalogues \cite{DESI:2025fxa},\footnote{\url{https://data.desi.lbl.gov/doc/releases/dr1/}.} both before and after density-field reconstruction, which we shall refer to as \emph{pre-} and \emph{post-reconstruction} catalogues, respectively. Reconstruction estimates the displacement field from the observed galaxy density and partially reverses the non-linear bulk flows that broaden and mildly shift the acoustic feature, with the aim of sharpening the BAO peak and improving the precision and accuracy of the recovered scale \cite{Eisenstein:2006recon, Seo:2008recon, Padmanabhan:2008recon, Seo:2010recon}. Reconstruction itself is performed after mapping the observed redshifts into comoving coordinates using a fiducial cosmology, and the post-reconstruction catalogues therefore inherit this intermediate model choice. However, extensive studies have shown that the recovered BAO scale is insensitive, at the precision relevant here, to reasonable variations of the fiducial cosmology adopted for reconstruction \cite{Sherwin:2018wbu, Carter:2019ulk}. The pre-reconstruction angular measurement is therefore the version of the analysis that is fully independent of this intermediate coordinate mapping, while the post-reconstruction measurement provides the appropriate like-for-like comparison with the official DESI DR1 3D BAO analysis, which is also performed after reconstruction. Therefore, we adopt the post-reconstruction measurements as our headline results and use them for the cosmological inference. We present these measurements in Section \ref{sec:res} for both pre- and post-reconstruction. 

\subsection{DESI DR1}

We measure the angular BAO signal using the DESI DR1 data release (DR1) large-scale structure catalogues. The tracers we study in this work include: Luminous Red Galaxies (LRGs), Emission Line Galaxies (ELGs) and Quasars (QSOs). We adopt equivalent redshift sample definitions as in the DESI clustering analyses. Full details on the target selection, survey properties and details of the pipeline to process the large-scale structure catalogue can be found in the relevant collaboration papers \cite{DESI:2022xcl,DESI:2022nlo,Myers:2022azg,DESI:2022gle, Raichoor:2022jab, Chaussidon:2022pqg}. Table \ref{tab:datatracers} summarises the relevant information for this work for each tracer population.

In addition to the data, we make use of two sets of mock suites for the pipeline validation and covariance estimation, respectively: 
\abacussummit mocks and EZmocks.

\begin{table}
    \centering
    \begin{tabular}{|l|c|c|c|}
      \hline
      Tracer   & $z$ range & $z_{\rm eff}$ & $N_{\rm tracer}$ \\
      \hline
      LRG1 & $0.4 - 0.6$ & 0.51 & 506,905 \\
      LRG2 & $0.6 - 0.8$ & 0.71 & 771,875 \\
      LRG3 & $0.8 - 1.1$ & 0.92 & 859,824 \\
      LRG3+ELG1 & $0.8 - 1.1$ & 0.93 & 1,876,164 \\
      ELG2 & $1.1 - 1.6$ & 1.32 & 1,415,687 \\
      QSO & $0.8 - 2.1$ & 1.49 & 856,652 \\
      \hline
    \end{tabular}
    \caption{We present a summary of the statistics for each of the tracers under study in this work. The redshift range represents the entire span for each tracer. We tabulate equivalent effective redshifts as in \cite{DESI:2024DR1}, as we use the same sample definitions and redshift edges.}
    \label{tab:datatracers}
\end{table}

\subsection{\abacussummit mocks}

The \abacussummit mocks are high-fidelity flagship mock catalogues \cite{Maksimova:2021ynf} built from large-volume, high-resolution cosmological $N$-body simulations. The entire suite of simulations consists of over 140 simulations exploring 97 different cosmologies. For this work, we make use of the 25 independent ``base'' simulations of the Planck 2018 $\Lambda {\rm CDM} $ cosmology (each with a box side of $2 \,\, h^{-1} {\rm Gpc}$). These simulations accurately model the non-linear evolution of the matter density field and populate dark matter haloes with DESI-like tracers, producing a realistic clustering scenario for which the pipeline of this work can be tested and validated on \cite{Yuan:2023ezi, Rocher:2023zyh}. We refer the reader to Appendix \ref{app:pipeval} for details on the pipeline validation.

An important observational artefact that should be accounted for when testing and validating the pipeline is the effect of fiber collisions. Due to the fact that each DESI fiber is mounted on a robotic positioner with a finite patrol region, and neighboring positioners must avoid mechanical collisions, close pairs of targets cannot always be observed simultaneously in a given tile. Thus, we make use of \abacus\ \altmtl mocks for pipeline validation, as they provide the most realistic treatment of this effect. The \altmtl procedure generates alternative merged target ledgers by varying the random subpriority values used by the DESI fiber-assignment algorithm and rerunning the fiber-assignment process. They are post-processed in a way that best mimics the DR1 large-scale clustering catalogues. Details on the fiber-assignment pipeline and the \altmtl procedure can be found in \cite{DESI:2024zvr, Bianchi:2024fnl}. The key takeaway for this work is that the \altmtl procedure reduces the assigned samples to approximately $20\%$ of the complete ELG catalogue, $56\%$ of the complete LRG catalogue, and $58\%$ of the complete QSO catalogue. 

\subsection{EZmocks}

The EZmocks \cite{Chuang:2014vfa} are approximate mock catalogues generated using the effective Zel'dovich approximation. They are designed to reproduce the large-scale clustering, number density, redshift distribution, and survey geometry of the DESI tracer samples at much lower computational cost than full $N$-body mocks. This makes it possible to generate a large number of independent realisations. In this work, we use the 1000 EZmocks generated for DESI DR1 to estimate the covariance matrix of the measured two-point angular correlation function. For each tracer and redshift snapshot, independent cubic boxes of $6 \,\, h^{-1} {\rm Gpc}$ side are generated and converted to galaxy fields using a bias model calibrated snapshot-by-snapshot to the two-point clustering of the DESI \abacus HOD mocks, which are based on the \abacussummit Planck 2018 $\Lambda{\rm CDM}$ cosmology. The boxes are then converted to the DESI sky geometry, and the different snapshots are concatenated into pseudo-lightcones before applying the DESI DR1 footprint and the fiber-assignment procedure.

The covariance matrix is presented later in Section \ref{sec:cov}. Similar to the \abacus mocks, we make use of the EZ Fast Fiber Assignment (FFA) mocks. Unlike the \abacus validation tests, which use the more realistic treatment \altmtl, the FFA procedure is an efficient approximation of the incompleteness of the DESI fiber assignment, allowing for the scalability of FFA to many mocks. The effect of FFA on the suppression of the number of tracers is similar to \altmtl. Details of the implementation of the FFA and \altmtl pipelines to the mocks and the differences between the two can be found in \cite{Bianchi:2024fnl}.

\section{Methodology}\label{sec:methods}

\subsection{Measuring $w(\theta)$}

We measure the two-point angular correlation function (2PACF) in thin tomographic redshift slices using the Landy-Szalay estimator \cite{LS1993ang},
\begin{equation}
    w(\theta) = \frac{DD(\theta) - 2DR(\theta) + RR(\theta)}{RR(\theta)},
\end{equation}
where $DD(\theta)$, $DR(\theta)$, and $RR(\theta)$ are the normalised data-data, data-random, and random-random pair counts, respectively, measured as a function of the angular separation $\theta$. The random catalogue follows the same angular footprint and selection function as the data, and therefore accounts for survey geometry in the estimator. The Landy-Szalay estimator is used because of its well-known minimum-variance properties for two-point clustering measurements in the presence of a finite survey window. Pair counts are computed using {\tt TreeCorr} \cite{Jarvis2015}, applying identical angular and redshift binning to both data and random catalogues. 

The measurements are performed in thin, non-overlapping redshift shells of width $\Delta z$, in which we measure only the angular auto-correlation function $w(\theta)$. Our fiducial redshift bin widths are $\Delta z = 0.02$ for the LRG1, LRG2, and LRG3+ELG1 samples, and $\Delta z = 0.05$ for the ELG2 and QSO samples. The corresponding angular bin widths are $\Delta\theta = 0.4^\circ$ for LRG1, LRG2, and LRG3+ELG1, and $\Delta\theta = 0.25^\circ$ for ELG2 and QSO. The remainder of this section motivates these fiducial choices. We note that the measurement pipeline was validated using mock catalogues and fixed before interpreting the DESI DR1 measurements.

The total transverse clustering information available for a given tracer is set by the survey footprint and the fitted range of $\theta$. The redshift binning determines how this information is redistributed among the shells and how much noise is injected into each slice. Thin shells are required to resolve the transverse BAO signal. In the limiting case of an infinitesimally narrow shell ($\Delta z \rightarrow 0$), the BAO feature corresponds to a single angular scale on the sky with no mixing of radial modes. In practice, the bin width must remain finite, and two competing effects set the fiducial value. Increasing $\Delta z$ raises the number density and pair statistics within each bin, but projects tracers spanning a wider radial interval onto a common redshift plane, broadening and shifting the apparent BAO feature. Decreasing $\Delta z$ suppresses this radial projection, but reduces the number of tracers per slice, thus raising the noise in $w(\theta)$, while lengthening the data vector. We therefore adopt the narrowest bins that retain measurable BAO information per tracer, consistent with earlier angular BAO analyses \cite{Sanchez:2010zg,Carvalho:2015ica,Menote:2021jaq}. The narrower binning choice is possible for the LRG samples because their higher number density supports robust angular clustering in thin shells, whereas the wider bins for ELG2 and QSO compensate for their lower number density to improve the probability of detecting the BAO feature at adequate significance. The projection behaviour is quantified in Figure 3 of \cite{Sanchez:2010zg}, where the shift of the fitted angular BAO scale is measured from the true angular BAO scale as a function of redshift bin width for different fiducial cosmologies. Although their analysis is done for a photometric survey, projection effects are of purely geometric origin and are agnostic to survey footprint and tracer sample. For our fiducial choices in $\Delta z$, their results imply a sub-percent shift in the BAO angular scale relative to the true value, which is negligible compared to our statistical uncertainties. The effect is milder still here, as spectroscopic redshifts define sharp radial slices without the photometric-redshift smoothing that broadens the effective slice.

Additionally, we account for the fact that the width of the BAO peak decreases as redshift increases, due to the reduced non-linear damping of the acoustic feature at higher redshifts. Thus, the angular binning has been chosen independently for each tracer so that the bin width is fine enough to sample the BAO peak with sufficient resolution, but not so fine that individual pair-count bins become dominated by shot noise. 

Finally, we note that because the redshift slices are narrow and non-overlapping, the auto-correlation in each slice provides the most direct measurement of the angular BAO scale associated with that interval, and we do not include cross-correlations between slices in the data vector. For non-overlapping thin redshift shells, the physical overlap between the tracer samples is small, so cross-correlations are expected to contain limited additional transverse BAO information relative to the auto-correlations. Such cross-bin correlations may nevertheless be sensitive to effects such as redshift-bin leakage, line-of-sight projection, or large-scale radial correlations. These are not the primary target of the present angular BAO analysis, which aims to test whether the angular BAO feature can be robustly recovered from individual thin shells as a first-stage validation of the angular BAO approach on DESI DR1 data, before extending the framework to a more complete tomographic analysis including cross-correlations. We defer such an extension, and the additional information it may carry, to future work with DESI DR2.

\subsection{Extracting the BAO signal}

We extract the BAO scale from the angular correlation function using a template-based algorithm. Template-based methods have been extensively used for BAO science \cite{Song:2024taz,Albornoz:2016lsx,DES:20242dbao,mena2026dark, DESI:2025DR2}. We first describe the computation of the template from which we construct our model $w(\theta)$ and, then, the BAO-fitting approach used in our analysis. We use the code \baofitwtheta, which is publicly available on \github,\footnote{\url{https://github.com/juanejo95/BAOfit_wtheta}.} and described in Appendix A of \cite{mena2026dark}.

\subsubsection{The BAO template}

The methodology we adopt to construct the template for the two-point angular correlation function employed in the BAO fits is as follows. We begin with the linear matter power spectrum, $P_{\rm m}^{\rm lin}(k)$, where $k$ denotes the wavenumber. At BAO scales, the dominant effect of non-linear evolution is the broadening of the BAO feature due to large-scale bulk flows. We model this effect by applying a Gaussian damping to the oscillatory (``wiggle'') component of the power spectrum, after separating it from the smooth broadband contribution (``no-wiggle''). Under these assumptions, the galaxy power spectrum in redshift space can be written as
\begin{gather}\label{eq:pkmu}
    P_{\rm g}(k,\mu,z)=\left[b(z)+f(z)\mu^2\right]^2\left[P_{\rm m}^{\rm w}(k)e^{-k^2\Sigma_{\rm tot}^2(\mu,z)}+P_{\rm m}^{\rm nw}(k)\right],
\end{gather}
where $P_{\rm m}^{\rm nw}$ describes the smooth shape of the linear matter power spectrum (the no-wiggle linear matter power spectrum) and all the BAO information is in $P_{\rm m}^{\rm w}(k)\equiv P_{\rm m}^{\rm lin}(k)-P_{\rm m}^{\rm nw}(k)$.\footnote{\baofitwtheta has the option to compute the template using only the no-wiggle component of the matter power spectrum, i.e., by removing the $P_{\rm m}^{\rm w}(k)$ term in \ref{eq:pkmu}. In our analysis, this option is used to estimate the significance of the BAO detection.} Both components are evaluated at $z=0$, and the redshift evolution of their amplitude is included through the growth factor $D(z)$ in Equation \ref{eq:wtheta_th}. The factor $\left(b+f\mu^2\right)^2$ accounts for linear-theory redshift-space distortions, where $\mu$ is the cosine of the angle of the wave vector $\boldsymbol{k}$ relative to the line of sight, $b$ is the linear galaxy bias and $f$ is the logarithmic derivative of the growth factor with respect to the scale factor. The no-wiggle component is modelled following the prescription described in Appendix D of \cite{brieden2022model}, using the implementation in {\sc cosmoprimo}.\footnote{\url{https://github.com/cosmodesi/cosmoprimo.git}.} The wiggle component is also computed with {\sc cosmoprimo}.

Following appendix A of \cite{ivanov2020cosmological}, the anisotropic damping factor, $\Sigma_{\rm tot}^2(\mu,z)$, can be expressed as
\begin{equation}\label{eq:Sigma_tot}
    \Sigma_{\rm tot}^2(\mu,z)=\mu^2\Sigma_\parallel^2(z)+(1-\mu^2)\Sigma_\perp^2(z)+f^2(z)\mu^2(\mu^2-1)\delta\Sigma^2(z).
\end{equation}
The parallel and perpendicular components of $\Sigma(z)$ are given by
\begin{equation}
    \Sigma_\parallel(z)=\left[1+f(z)\right]\Sigma(z),\quad\Sigma_\perp(z)=\Sigma(z),
\end{equation}
and the redshift evolution is determined by the growth factor $D(z)$:
\begin{equation}
    \begin{dcases}
        \Sigma(z)=D(z)\Sigma_0,\\
        \delta\Sigma(z)=D(z)\delta\Sigma_0.
    \end{dcases}
\end{equation}
The $(1+f)$ factor in $\Sigma_\parallel(z)$ is due to the Kaiser formula. At $z=0$, we find
\begin{equation}
    \begin{dcases}
        \Sigma_0^2=\frac{1}{6\pi^2}\int_0^{k_s}\dd{q}P_{\rm m}^{\rm nw}(q)\left[1-j_0(q\ell_{\rm BAO})+2j_2(q\ell_{\rm BAO})\right],\\
        \delta\Sigma_0^2=\frac{1}{2\pi^2}\int_0^{k_s}\dd{q}P_{\rm m}^{\rm nw}(q)j_2(q\ell_{\rm BAO}),
    \end{dcases}
\end{equation}
where $j_n$ are the spherical Bessel functions of order $n$, while $\ell_{\rm BAO}$ is the wavelength of the BAO. Following \cite{ivanov2020cosmological}, we choose $k_s=0.2$ $h$/Mpc and $\ell_{\rm BAO}=110$ Mpc/$h$ as reference values.

Since we restrict our analysis to autocorrelations within a given tomographic bin $i$, we approximate the redshift evolution of the power spectrum by evaluating Equation \ref{eq:pkmu} at the average redshift of that tomographic bin, $\bar{z}^i$.\footnote{We model the redshift distributions $\phi(z)$ as top-hat functions within each redshift slice. Therefore, the average redshift $\bar{z}^i$ is given by the center of the slice.} This approximation is justified by the narrow width of the redshift bins considered. In the following, we therefore define
\begin{equation}
    P_{\rm g}^i(k,\mu) \equiv P_{\rm g}(k,\mu,\bar{z}^i).
\end{equation}
We then decompose the anisotropic power spectrum into Legendre multipoles, $\mathcal{L}_\ell$,
\begin{equation}
    P_{\rm g,\ell}^i(k)=\frac{2\ell+1}{2}\int_{-1}^1\dd{\mu} P_{\rm g}^i(k,\mu)\mathcal{L}_\ell(\mu),
\end{equation}
which isolate the angular dependence induced by redshift-space distortions and the anisotropic BAO damping. The corresponding multipoles of the two-point correlation function are obtained via a Fourier transform,
\begin{equation}
    \xi_{\rm g,\ell}^i(s)={\rm i}^\ell\int_0^\infty \dd{k}j_\ell(ks)\frac{k^2}{2\pi^2}P_{\rm g,\ell}^i(k),
\end{equation}
where i is the imaginary unit, and the full anisotropic correlation function can be reconstructed as
\begin{equation}
    \xi_{\rm g}^i(s,\mu)=\sum_{\ell}\xi_{\rm g,\ell}^i(s)\mathcal{L}_\ell(\mu),
\end{equation}
where in practice, we use $\ell = 0, 2, 4$.  To relate the previous expression to observable quantities, we express the comoving separation $s$ and the cosine of the line-of-sight angle $\mu$ in terms of the redshifts and angular separation of a pair of galaxies, $(z_1, z_2, \theta)$,
\begin{gather}
    \xi_{\rm g}^i(z_1,z_2,\theta)=\xi_{\rm g}^i\left(s(z_1,z_2,\theta),\mu(z_1,z_2,\theta)\right)\nonumber\\
    =\sum_\ell\xi_{\rm g,\ell}^i\left(\sqrt{\chi_1^2+\chi_2^2-2\chi_1\chi_2\cos\theta}\right)\mathcal{L}_\ell\left(\frac{\chi_1-\chi_2}{\sqrt{\chi_1^2+\chi_2^2-2\chi_1\chi_2\cos\theta}}\right),
\end{gather}
where $\chi_j = \chi(z_j)$ denotes the comoving distance to each galaxy. Finally, the angular correlation function of the $i$-th bin is obtained by projecting $\xi_{\rm g}^i(z_1,z_2,\theta)$ along the line of sight using the corresponding redshift distribution, $\phi^i(z)$,
\begin{equation}\label{eq:wtheta_th}
    w^i(\theta)=\int \dd{z_1}D(z_1)\phi^i(z_1)\int \dd{z_2} D(z_2)\phi^i(z_2)\xi_{\rm g}^i(z_1,z_2,\theta),
\end{equation}
where the growth factor $D(z)$ accounts for the residual redshift evolution of the matter clustering. We compute the template $w(\theta)$ over the range $0.001^\circ \leq \theta \leq 179.5^\circ$ using 1000 logarithmically spaced bins. The cosmology used for the template throughout this paper, which we refer to as ``fiducial'' or ``template'' cosmology interchangeably, is Planck 2018.

\subsubsection{Linear galaxy bias}

Inspecting the galaxy power spectrum in Equation \ref{eq:pkmu}, we note that the dependence on the linear galaxy bias arises entirely from the Kaiser factor, $(b + f\mu^2)^2$. Expanding this term, the power spectrum can be written as a sum of contributions proportional to $b^2$, $bf$ and $f^2$, where the last term is independent of the galaxy bias.

Since the angular correlation function defined in Equation \ref{eq:wtheta_th} depends linearly on the galaxy power spectrum, this decomposition carries over directly to configuration space. As a result, the angular correlation function for the $i$-th tomographic bin can be expressed as
\begin{equation}\label{eq:w_bb_bf_ff}
    w^i(\theta)=b^2w_{b^2}^i(\theta)+bw_{bf}^i(\theta)+w_{f^2}^i(\theta),
\end{equation}
where each term corresponds to the contribution associated with a given power of the bias. In practice, we compute and store these three components ($w_{b^2}$, $w_{bf}$ and $w_{f^2}$) separately. This decomposition allows us to determine the linear galaxy bias from the data $w(\theta)$ (we do so on small scales, where the signal is most sensitive to $b$). We then use these measured linear galaxy bias values to produce the template for the BAO fits via Equation \ref{eq:w_bb_bf_ff}. This procedure ensures an accurate normalization of the clustering signal when comparing the template to the data (for both mocks and data).

In the next section, we describe the BAO modelling used in this work, where we introduce a parameter $A^i$ that rescales the overall clustering amplitude. Since, as we just explained, the template is normalised using the bias estimated from the data, $A^i$ is expected to be close to unity and absorb any residual mismatch in the amplitude. On the other hand, $f$, together with $\Sigma_{\rm tot}^2(\mu,z)$ in Equation \ref{eq:pkmu}, is evaluated at $\bar{z}^i$, the effective redshift of tomographic bin $i$. The impact of these modelling choices is partly mitigated by allowing other coefficients, such as $A^i$ and $a_j^i$, to vary. However, this is not fully equivalent to treating $b$ and $f$ as independent free parameters.

\subsection{The BAO fitting procedure}\label{ssec:baofitting}

With the template construction outlined in the previous section, we can build a model to fit the BAO signal in the $i-$th redshift bin. In what follows, the tracer bin is kept fixed, so the index $i$ labels the redshift sub-bin within that tracer sample, and $z_i$ denotes its effective redshift, while $z_{\rm eff}$ denotes the effective redshift of the full tracer sample. The model is given by
\begin{equation}\label{eq:wtheta_model} 
    w_{\rm model}^{i}(\theta,{\bf \Theta}) = A^i\, w_{\rm temp}^i(\alpha_{\perp} \theta) + \sum_j a_j^i\, \theta^{-j}, 
\end{equation}
where $w_{\rm temp}^i(\theta)$ is the theoretical BAO template constructed in the fiducial cosmology, computed using Equation \ref{eq:wtheta_th}. ${\bf \Theta}$ collects the free parameters of the model and is specified in Equation \ref{eq:theta_joint} below. The parameter $A^i$ rescales the overall clustering amplitude and the coefficients $a_j^i$ model the residual differences in the broadband contribution, with $j$ denoting the power-law index for the polynomial in $\theta$. In our work, we use $j = 0, 1$. This choice has been extensively tested on mocks for all tracers, and we direct the reader to Appendix \ref{app:pipeval} for more information on this test. In brief, retaining only two broadband terms provides the model with enough freedom to capture the broadband signal while avoiding excessive flexibility that could absorb the BAO signal.

The parameter $\alpha_{\perp}$ is the transverse dilation parameter which, in the context of the fitting procedure, shifts the angular position of the BAO feature relative to the reference template.
Consequently, a BAO feature located in the template at $\theta_{\rm BAO,temp}(z_i)$ appears in the model at the angle satisfying $\alpha_{\perp}\theta_{\rm BAO}(z_i) = \theta_{\rm BAO,temp}(z_i)$, so that
\begin{equation}\label{eq:alpha_thetafitting}
    \theta_{\rm BAO}(z_i)= \frac{\theta_{\rm BAO,temp}(z_i)}{\alpha_\perp}.
\end{equation}
Thus, it is clear that a value of $\alpha_{\perp} = 1 $ corresponds to no shift relative to the template cosmology. Equation \ref{eq:alpha_thetafitting} defines $\alpha_{\perp}$ within the context of the fitting procedure, but equivalently, we can define this in terms of the standard geometrical relation. Since the angular BAO scale is given by
\begin{equation}\label{eq:thetabao}
    \theta_{\rm BAO}(z) = \frac{r_{\rm d}}{D_M(z)},
\end{equation}
we can rewrite $\alpha_{\perp}$ in relation to the transverse distance through
\begin{equation}
    \alpha_\perp
    =\frac{D_M(z_{\rm eff})/r_{\rm d} }{D_M^{\rm temp}(z_{\rm eff})/r_{\rm d}^{\rm temp}},
    \label{eq:alpha_perp_definition}
\end{equation}
where $z_{\rm eff}$ is the effective redshift of the tracer sample, $D_M$ is the transverse comoving distance and $r_{\rm d}$ is the sound horizon at the drag epoch. Sharing a single $\alpha_\perp$ across the redshift bins of a tracer assumes that this ratio varies negligibly over the redshift range of the sample, and we therefore quote it at $z_{\rm eff}$.

Equation \ref{eq:alpha_perp_definition} is what makes the fitted dilation parameter physically useful. For each redshift bin, we fit the measured $w(\theta)$ against the model (marginalising over the amplitude and broadband nuisance terms) to obtain a posterior on $\alpha_\perp^i$. The template quantities $D_M^{\rm temp}(z_i)$ and $r_{\rm d}^{\rm temp}$ are fixed and are set to be the $\Lambda {\rm CDM}$ Planck 2018 cosmology. Inverting Equation~\ref{eq:alpha_perp_definition} converts each fitted $\alpha_\perp^i$ directly into a measurement of the transverse distance-to-sound-horizon ratio,
\begin{equation}
    \frac{D_M(z_{\rm eff})}{r_{\rm d}} = \alpha_\perp \, \frac{D_M^{\rm temp}(z_{\rm eff})}{r_{\rm d}^{\rm temp}}.
    \label{eq:DM_rd_from_alpha}
\end{equation}
The dimensionless parameter $\alpha_\perp$ therefore encodes the departure of the measured BAO scale from that of the template, and Equation \ref{eq:DM_rd_from_alpha} maps this departure onto the physical observable $D_M(z_{\rm eff})/r_{\rm d}$. Repeating the fit for each tracer sample yields $D_M/r_{\rm d}$ at five effective redshifts, which constitutes the data vector we use to constrain cosmological parameters in Section \ref{sec:cosmoinf}. The template cosmology enters only as a fixed reference. To first order, a different choice of $D_M^{\rm temp}(z_{\rm eff})/r_{\rm d}^{\rm temp}$ rescales the recovered $\alpha_\perp$ by the compensating factor, leaving the inferred $D_M(z_{\rm eff})/r_{\rm d}$ unchanged.

We crop the data vector within $\Delta \theta_{\rm fit}$ centred at the BAO peak, $\theta_{\rm BAO}(z) = r_{\rm d}/D_M(z)$, for each individual redshift bin. This keeps the length of the data vector under control and improves the stability of the covariance matrix inversion. Analogously to the choice of broadband terms, the choice of $\Delta \theta_{\rm fit}$ for each tracer sample has been extensively tested for the LRG case. We found that the most sensible choice is $\Delta  \theta_{\rm fit} = 3^\circ$ from the BAO peak for LRG1, LRG2 and LRG3+ELG1, and $\Delta  \theta_{\rm fit} = 1.5^\circ$ for ELG2 and QSO. For the latter samples, the narrower window is mainly dictated by the position of the BAO peak, since a wider interval would extend too close to, or beyond, the lower boundary of the measured range. As with the redshift and angular binning, these fitting ranges were fixed from mock-based validation prior to analysis of the DESI data.
We denote by $N_{\theta}^i$ the number of angular bins retained in redshift bin $i$ after this cut, and index them by $k = 1,\ldots,N_{\theta}^i$ and by $\theta_k^{i}$ for the angular separations for each redshift bin. The measured correlation function in redshift bin $i$ and angular bin $k$ is written $w^i(\theta_k^{i})$.

The measurements are stacked into a single data vector $\wvec$ flattened along the redshift and retained angular bins,
\begin{equation}\label{eq:datavector}
    \wvec =
    \Big(  w^1(\theta^1_1),\ldots,w^1(\theta^1_{N_\theta^1}),
        \ldots,
        w^{N_z}(\theta^{N_z}_1),\ldots,w^{N_z}(\theta^{N_z}_{N_\theta^{N_z}})
    \Big)^{\! \top},
\end{equation}
whose components we label $w_a$, with the flattened index $a \equiv a(i,k)$ running over $a = 1,\ldots,N_{\rm data}$. $N_z$ denotes the number of redshift bins and 
\begin{equation}\label{eq:Ndata}
    N_{\rm data} = \sum_{i=1}^{N_z} N_{\theta}^i.
\end{equation}
Note that $N_{\theta}^i$ varies between redshift shells for the samples spanning the widest redshift range, since the fitting window is centred on the redshift-dependent BAO peak.

To quantify the goodness-of-fit, we can construct the residual $\Delta {\bf w(\theta,\bf{\Theta})}$ between the data and model,
\begin{equation}
    \Delta \wvec(\bf{\Theta})= {\bf w_{\rm data}} - {\bf w_{\rm model}(\bf{\Theta})},
\end{equation}
with components $\Delta w_a = w_{{\rm data}, a} - w_{{\rm model}, a} $, ordered as in Equation \ref{eq:datavector}.
We use the Markov Chain Monte Carlo (MCMC) Metropolis sampler {\tt cobaya} \cite{Torrado:2020dgo} and assume a Gaussian likelihood, with a covariance matrix $\Cmat$, of the form
\begin{equation}\label{eq:likelihood}
    -2\ln\mathcal{L}(\boldsymbol{\Theta})
    =\Delta \wvec({\bf \Theta})^T\,\Cmat^{-1}\, \Delta \wvec({\bf \Theta}).
\end{equation}

The 2PACFs of all redshift bins within a tracer sample are fitted simultaneously, with a single $\alpha_\perp$ parameter shared across all bins. The parameter vector is, therefore,
\begin{equation}\label{eq:theta_joint}
    \boldsymbol{\Theta}= \{ \alpha_\perp, A^1, a_0^1 ,a_1^1,\ldots , A^{N_z}, a_0^{N_z} ,a_1^{N_z}\},
\end{equation}
giving a total of $1 + 3N_z$ free parameters. This is our default setup for the results in this paper: a single $\alpha_\perp$ per tracer sample (i.e., LRG1, LRG2, LRG3+ELG1, ELG2 and QSO). We also consider individual-bin fits, in which all redshift bins but bin $i$ are removed from the fit. In that case, the parameter vector reduces to
\begin{equation}\label{eq:theta_indep}
    \boldsymbol{\Theta}= \{ \alpha_\perp, A^i, a_0^i ,a_1^i\},
\end{equation}
with a total of 4 free parameters, and the fitted $\alpha_\perp$ is interpreted at $z_i$ rather than $z_{\rm eff}$ in Equation \ref{eq:DM_rd_from_alpha}. We denote the result of the fit to bin $i$ by $\alpha_\perp^i$, and present these results in Appendix \ref{app:individualfit}.

\section{Covariance matrix}\label{sec:cov}

\begin{table}
    \centering
    \begin{tabular}{lcccc}
        \toprule
        Tracer & $N_z$ & $N_{\rm data}$  & $\psi$ \\
        \midrule
        LRG1        & 10 & 152   & 0.847 \\
        LRG2        & 10 & 144  & 0.855 \\
        LRG3+ELG1   & 15 & 190  & 0.809 \\
        ELG2        & 10 & 200  & 0.799 \\
        QSO         & 26 & 511 & 0.487 \\
        \bottomrule
    \end{tabular}
    \caption{Length of the fitted data vector and the corresponding Hartlap correction, Equation \ref{eq:hartlap}, for each tracer sample. The QSO sample has by far the longest data vector and hence the most severe correction. Its robustness is examined in Section~\ref{sec:qso_cov} and Appendix~\ref{app:offblock}.}
    \label{tab:hartlap}
\end{table}

For the BAO fitting pipeline, we require a covariance matrix for the final data vector. In this work, we use a mock-based covariance computed from $N_{\rm mocks}= 1000$ EZ FFA mock catalogues as the baseline. The choice is motivated by the fact that a mock-based covariance by construction includes the effects of survey geometry, assignment incompleteness, shot-noise, and estimator-level effects, none of which are straightforward to model analytically for a tomographic angular measurement. 
Let $\wvec^{(m)}$ denote the data vector measured on mock realisation $m = 1,\ldots,N_{\rm mocks}$, constructed exactly as in Equation~\ref{eq:datavector} and with components $w_a^{(m)}$. Each mock is analysed through a measurement pipeline identical to the one applied to the data. We note that we apply the angular cut that defines $N_{\theta}^i$ to the covariance matrix before inversion. 
Writing the mean data vector across all mocks as 
\begin{equation}\label{eq:meanw}
    \bar{\wvec} = \frac{1}{N_{\rm mocks}}\sum_{m=1}^{N_{\rm mocks}} \wvec^{(m)},
\end{equation}
the elements of the mock covariance matrix $\Cmat$ are
\begin{equation}\label{eq:covariance}
    C_{ab} = \frac{1}{N_{\rm mocks} - 1}\sum_{m=1}^{N_{\rm mocks}}
      \left(w_a^{(m)} - \bar{w}_a\right)
      \left(w_b^{(m)} - \bar{w}_b\right),
\end{equation}
with $a,b = 1,\ldots,N_{\rm data}$. This is the matrix $\Cmat$ entering the likelihood in Equation \ref{eq:likelihood}.

Since the covariance is estimated from a finite ensemble of mocks, the inverse of the sample covariance matrix is a biased estimator of the true precision matrix \cite{Hartlap:2006kj}. We correct for this bias by rescaling the inverted covariance matrix ${\sf C}^{-1}$ by the Hartlap factor,
\begin{equation}\label{eq:hartlap}
    \psi = \frac{N_{\rm mocks} - N_{\rm data} - 2}{N_{\rm mocks} - 1},
\end{equation}
so that $\Cmat^{-1} \rightarrow \psi\,\Cmat^{-1}$. Table \ref{tab:hartlap} tabulates $N_{\rm data}$ and $\psi$ for each tracer sample. The correction is modest for the LRG and ELG samples, but substantial for the QSO sample, whose 26 tomographic shells give $N_{\rm data} = 511$ and hence $\psi = 0.49$, corresponding to an inflation of the parameter uncertainties by a factor of 1.43.

\begin{figure}
    \centering
    \includegraphics[width=\linewidth]{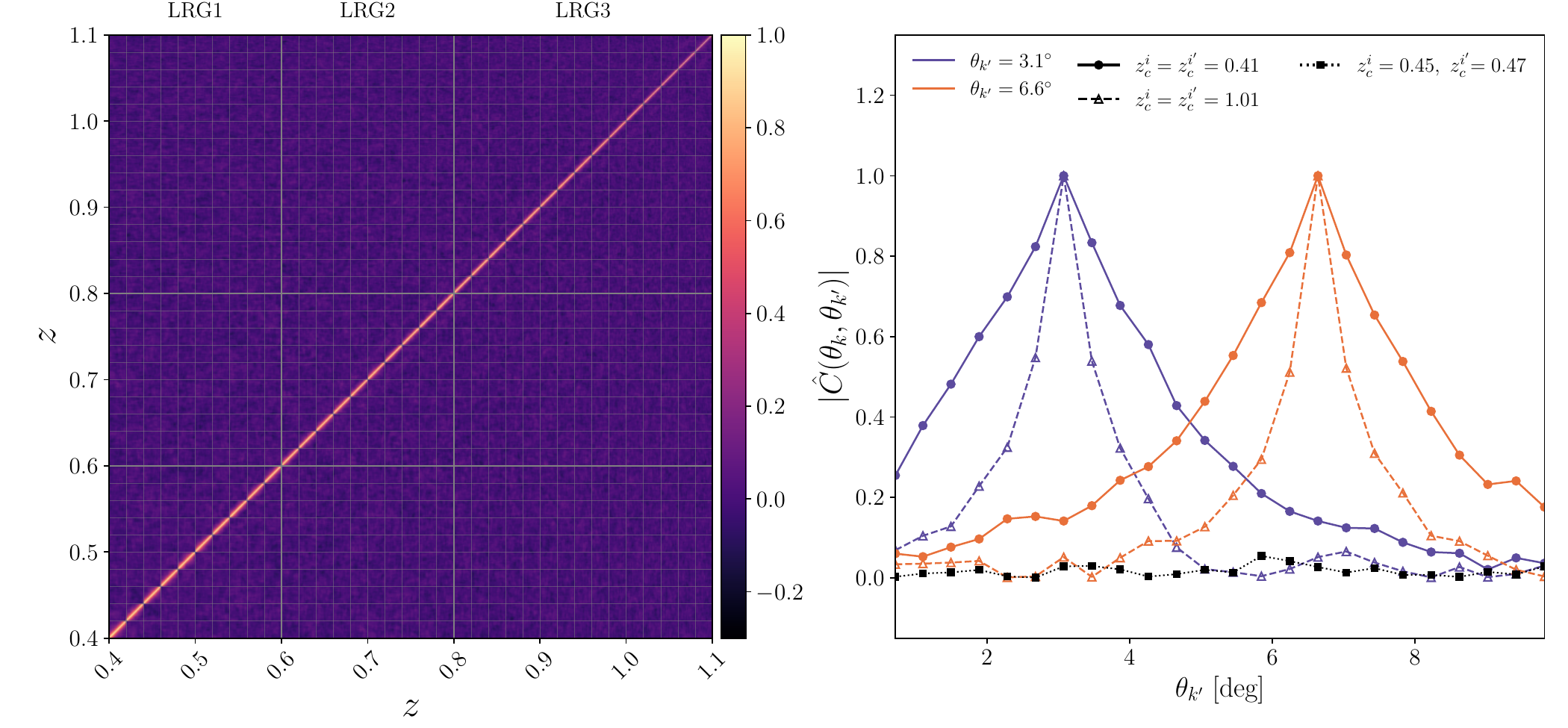}
    \caption{{\it Left panel}: Correlation matrix $\Ccorr$ computed from 1000 EZ FFA mocks for the LRG sample. The grey delimiters indicate the redshift ranges of the subtracers: LRG1, LRG2 and LRG3. {\it Right panel}: The individual rows of the absolute of the correlation matrix $\Ccorr$ for the LRG sample at two fixed $\theta_k$ as a function of $\theta_{k'}$, fixed within a given redshift shell. The dashed line and transparent triangle points are the behaviour at high redshift within the same redshift shell, and the solid lines and circular points denote the behaviour at low redshift within the same redshift shell. The black square and dotted lines denote the absolute average correlation in adjacent redshift shells.}
    \label{fig:cov}
\end{figure}

It is also convenient to define the correlation matrix, $\Ccorr$, with the elements given by
\begin{equation}
    \hat{C}_{ab} = \frac{C_{ab}}{\sqrt{C_{aa}{C_{bb}}}},
\end{equation}
which we use to visualise the structure of $\Cmat$ in  Figure \ref{fig:cov}. We show the correlation matrix $\mathsf{\hat{C}}$ of the LRG data vector across $0.4 < z < 1.1$ as a representative example. The remaining tracers display the same qualitative structure. 

The left panel of Figure \ref{fig:cov} shows $\Ccorr$ for the full flattened data vector, with both axes running over the index $a$ and labelled by the central redshift of the corresponding shell. A clear block-diagonal structure is present. Each diagonal block corresponds to a single redshift shell, while correlations between different shells are strongly suppressed. Within each diagonal block, neighbouring angular bins are strongly correlated, as is characteristic of configuration-space measurements, in which each angular separation receives contributions from an overlapping range of harmonic modes. These terms are sourced by sample variance. The large-scale modes shared within each finite-volume realisation correlate adjacent $\theta$ bins, with the correlation decaying with angular separation. 

The right panel makes the structure explicit by showing individual rows of $\Ccorr$, plotted as $|\hat{C}(\theta_k,\theta_{k'})|$ for two fixed $\theta_k$ as a function of $\theta_{k'}$. We plot the behaviour at both high and low redshifts, in solid circles and transparent triangles, respectively. Within each redshift shell, the correlation peaks at $\theta_{k'} = \theta_k$ and decays approximately symmetrically on either side, falling to $\hat{C} \simeq 0.83$ for immediately adjacent bins. The decay is more rapid in the redshift bin with the highest redshift, $z_c^i = 1.01$, than in $z_c^i = 0.41$. The LRG number density falls steeply beyond $z \sim 0.8$, so the shot noise, which is diagonal in $\theta$, grows relative to the sample variance contribution responsible for the off-diagonal terms and suppresses the off-diagonal elements of $\Ccorr$. Rows drawn from adjacent redshift shells ($z_c^{i} = 0.45, z_c^{i} = 0.47$) have negligible correlation. Their residual amplitude, $|\hat{C}| \sim 0.03$, matches the noise floor expected from a finite ensemble of $N_{\rm mock} = 1000$ realisations, $1/\sqrt{N_{\rm mock}} \simeq 0.03$, and we therefore attribute it to sampling noise rather than to a physical correlation.

\subsection{Robustness of the QSO covariance}\label{sec:qso_cov}

The Hartlap correction is largest for the QSO sample, where $\psi = 0.49$ inflates the parameter uncertainties by $43\%$. A natural way to reduce it is to exploit the block-diagonal structure of the covariance. We decompose the mock covariance as $\Cmat = {\sf D} + {\sf N}$, where ${\sf D}$ contains the diagonal blocks associated with each redshift shell and ${\sf N}$ the off-diagonal part connecting distinct shells. The elements of ${\sf N}$ are consistent with sampling noise, with an root-mean-squared (rms) correlation of $\sim 0.03$ that matches the noise floor $1/\sqrt{N_{\rm mocks}}$ discussed above. Setting ${\sf N}$ to zero makes the covariance block diagonal, so that the inversion factorises into independent inversions of each diagonal block and the Hartlap correction applies with the block dimension in place of $N_{\rm data}$, giving $\psi_i \simeq 0.98$. Although this removes almost the entire correction, we found that this does not change $\alpha_{\perp}$. Once each covariance is corrected with its own Hartlap factor, the full and block-diagonal covariances yield the same constraint on $\alpha_\perp$. The large Hartlap correction for the QSO sample therefore reflects the removal of spurious information contained in ${\sf N}$ rather than a loss of genuine constraining power, and we retain the full mock covariance with the standard Hartlap correction throughout. We summarise the tests supporting this conclusion below, and present them in full in Appendix \ref{app:offblock}. 

Although the elements of ${\sf N}$ are individually small, their collective effect on the precision matrix is nevertheless substantial. This is because matrix inversion couples every element of the covariance, so discarding ${\sf N}$ changes the inverse throughout. We found that within the diagonal blocks ${\sf D}^{-1}$ differs from $\Cmat^{-1}$ by a median fractional amount of $-0.62$. A large difference between the two precision matrices $\Cmat^{-1}$ and ${\sf D}^{-1}$ does not necessarily imply a large difference in the inferred parameters. The likelihood depends on the precision matrix only through its projection onto the directions in data space in which the model parameters can actually change. For the QSO, with 511 data points and 79 modelling parameters (3 broadband terms for each of the 26 redshift bins, plus $\alpha_{\perp}$), these directions span only a small subspace. The differences in the precision matrix beyond these directions have no effect on the inference. We explore this projection quantitatively following a similar approach in Section 5 of \cite{NovellMasot:2023gmj}, and find two effects that we summarise here. First, ${\sf N}$ has a negligible impact on the recovered BAO scale, shifting the best-fit $\alpha_\perp$ by $0.25 \sigma$ of its statistical uncertainty. Second, ${\sf N}$ does increase the apparent precision on $\alpha_\perp$, with the full covariance yielding a Fisher uncertainty $\sim 30\%$ smaller than the block-diagonal one before any Hartlap correction. Once each covariance is corrected with the Hartlap factor, we find that using either $\Cmat$ or ${\sf D}$ yields the same uncertainty on $\alpha_\perp$ to within a few per cent. Both prescriptions also recover the input cosmology when fitted to the mean of the 25 \abacus\ \altmtl mocks that are consistent with one another. 

The fact that ${\sf N}$ is consistent with noise is also what we expect physically for this sample. Distinct redshift shells contain different quasars, so the shot-noise contribution to the covariance is confined to the diagonal blocks, and shells can only be correlated through cosmic variance from large-scale modes that they share. For the QSO sample, this contribution is strongly suppressed. Its number density is low enough \cite{DESI:2024cosmo} that the measurement is dominated by shot noise rather than cosmic variance. For this reason, the DESI DR1 3D BAO analysis also treated the QSO measurement as effectively independent of the LRG and ELG measurements that overlap with it in redshift \cite{DESI:2024cosmo}. We therefore expect the true cross-shell covariance of the QSO $w(\theta)$ to be small, and the off-diagonal elements measured from the mocks to be dominated by sampling noise, in agreement with what we find.

A related question is whether correlations between different tracer samples need to be included when the measurements are combined. The only samples that overlap in redshift are the QSO sample and the LRG3 sample at $0.8 < z < 1.1$, and the QSO sample and the ELG2 sample at $1.1 < z < 1.6$. The LRG3 and ELG1 samples also overlap at $0.8 < z < 1.1$, but we combine them into a single LRG3+ELG1 sample, so that no correlation between them needs to be modelled. The correlation between QSO and the other tracers is expected to be small, since the low number density of the QSO sample means that its measurement is dominated by shot noise, which is uncorrelated between different sets of objects. The DESI DR1 3D BAO analysis estimated this correlation to be below $0.1$ in both overlapping ranges and treated the QSO measurement as independent of the LRG and ELG measurements \cite{DESI:2024cosmo}. We follow the same approach and treat the measurements from different samples as independent in the cosmological inference of Section \ref{sec:cosmoinf}.

\section{Results}\label{sec:res}

\begin{figure}
    \centering
    \hspace*{-2cm}
    \includegraphics[width=\linewidth]{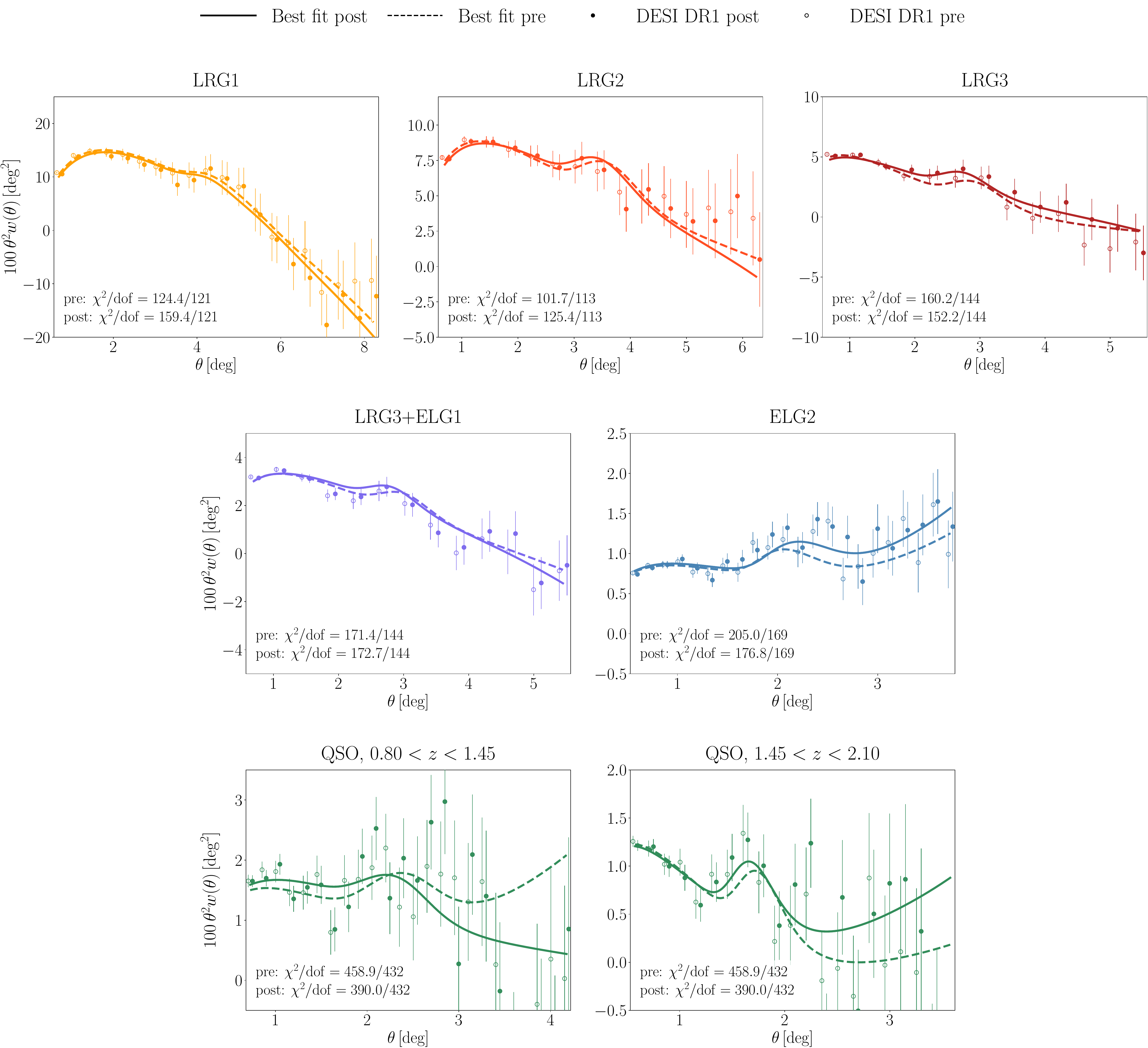}
    \hspace*{-2cm}
    \caption{Measurements of the two-point angular correlation function $w(\theta)$ for each tracer averaged over the thin redshift bins comprising that tracer sample for visualisation purposes (the fits themselves are per redshift bin; see Section \ref{ssec:baofitting}). We make the BAO peak more visible by plotting the quantity $100\,\theta^2\,w(\theta)$. Here we show the full measurement range in $\theta$ for each tracer. The range of the best-fit model here is the union of the angular bins retained for each redshift bin. The solid and transparent points indicate the post- and pre-reconstruction measurements, respectively. Solid and dashed lines indicate the corresponding post- and pre-reconstruction best-fit theoretical model. For visualisation purposes, the QSO sample is split into two redshift panels, and we minimally displaced the post- and pre- reconstruction points horizontally.}
    \label{fig:measurements}
\end{figure}

\subsection{$w(\theta)$ measurements and best-fit} 

We compute the 2PACFs and report them in Figure \ref{fig:measurements} along with their best-fit theoretical models (see Equation \ref{eq:wtheta_model}) for each tracer. We plot the quantity $100 \times \theta ^2 \times w(\theta)$ to visually enhance the angular BAO feature. For compactness and direct comparison with the DESI DR1 clustering measurements, we present in each panel the average of the data points and the best-fit model over the individual thin redshift bins that make up the tracer bin. We stress that these bin-averaged 2PACFs serve visualization purposes only, and that the BAO fits are always run by concatenating the $w(\theta)$ of the different redshift bins constituting the tracer sample as explained in Section \ref{ssec:baofitting} and not averaging them. For Figure \ref{fig:measurements}, we split the QSO sample into two redshift ranges because it spans a considerably broader redshift interval than the other tracers. Averaging over the full QSO range would superpose angular BAO features at systematically different angular scales, which would produce an artificially broadened or multi-peaked feature. 

There are several notable qualitative features to point out in Figure \ref{fig:measurements}. First, we are reassured that the angular BAO peak shifts towards smaller angles as the redshift increases, as expected given Equation \ref{eq:thetabao}. The measurements are also visibly noisy at large angular separations where the statistical uncertainties become large compared to the amplitude of the BAO. This is expected at those scales as we are sensitive to cosmic variance and survey mask geometry. There are essentially fewer independent large-scale modes contributing to the measurement at those scales. This is most apparent for the QSO samples, for which there is the additional caveat that we have a relatively lower number density compared to other tracers, as highlighted in Table \ref{tab:datatracers}. This results in noisier estimates of $w(\theta)$ in the individual redshift slices. 

We note that the best-fit model does not span the whole measured range. As described in Section \ref{ssec:baofitting}, we fit the data only within a $\Delta\theta$-wide interval around the expected position of the BAO peak, in order to ensure the stability of the covariance matrix inversion. Hence the solid line, which denotes the best-fit model averaged over the tracer bin, spans the union of the relevant theta range across all the redshift shells that make up the tracer. The measurements outside this range are not shown and they do not enter the quoted fits or the corresponding $\chi^2$ values. The $\chi^2$ reported in each panel is the minimum found with the \texttt{iminuit} minimiser over the fitted range.

Overall, the data points are well-described by the best-fit model across different tracer samples, with the quality of fit varying per tracer. The solid curves and points denote the post-reconstruction results, while the dotted and transparent points are the pre-reconstruction results. Contrary to the 3D BAO DESI DR1 measurements, reconstruction in the case of 2D BAO does not uniformly improve the fit across samples. The agreement improves for QSO. It lowers $\chi^2/{\rm dof}$ from $1.06$ to $0.90$, and for the ELG2 going from $1.21$ to $1.05$. There is a milder improvement for LRG3, and it leaves LRG3+ELG1 unchanged. For the other tracers, reconstruction seems to pull the $\chi^2/{\rm dof}$ values further from unity with respect to pre-reconstruction. For LRG1 and LRG2, the post-reconstruction fit is worse than the pre-reconstruction one, with LRG1 rising from $1.03$ to $1.32$. 

We tabulate the summary of the BAO-fitting results, along with the best-fit $\chi^2$, in Table \ref{tab:alpha_summary}. As shown in the Table, we find that the impact of reconstruction on the precision of the angular measurement is more modest and not uniform across tracers. The uncertainty decreases by $25\%$ for QSO and $11\%$ for LRG2, but increases by $7\%$ for LRG1 and $20\%$ for ELG2, with a mean improvement of only $\sim 4\%$ across the six samples. Even in 3D BAO analyses the gain from reconstruction is a statistical statement that holds on average for realisations but not for each individual realisation (See Figure 15 of \citep{Ross:2013vla}). Our mocks show similar behaviour. The mean posterior uncertainty decreases after reconstruction for the LRG and ELG2 samples but increases by $7\%$ for QSO and the DESI DR1 data values are within the range spanned by the mocks (Figure \ref{fig:sigmaalphamocks}). The angular statistic dilutes the expected gain in precision, since reconstruction sharpens the 3D density field while the angular correlation function retains only its projection onto the sky, and part of the recovered sharpening is washed out by the radial integration within each tomographic shell. However, the benefit of reconstruction is clearer in the recovery of the BAO scale itself. Across the mocks, the scatter of the recovered $\alpha_\perp$ decreases for every tracer, by up to $\sim 30\%$ for LRG2 and LRG3, and the mean recovered value moves closer to unity for every tracer (Figure \ref{fig:alphamocks} and Table \ref{tab:mock_alpha_summary} in Appendix \ref{app:pipeval}).

\begin{table}
    \centering
    \begin{tabular}{lccccc}
        \hline
        Tracer & $z_{\rm eff}$ & $\alpha_{\rm mean}$ & $\sigma_\alpha$ & $\alpha_{\rm bf}$  & $\chi^2_{\rm bf} / \mathrm{dof}$ \\
        \hline
        \multicolumn{6}{l}{\textbf{PRE}} \\
        \hline
        LRG1 & 0.51 & 0.9703 & 0.0259 & 0.9619 & 124.44 / 121 \\
        LRG2 & 0.71 & 0.9540 & 0.0290 & 0.9471 & 101.72 / 113 \\
        LRG3 & 0.92 & 0.9446 & 0.0272 & 0.9317 &  160.24 / 144  \\
        LRG3+ELG1 & 0.93 &  0.9321 & 0.0259 & 0.9244  & 171.39 / 144  \\
        ELG2 & 1.32 & 0.9943 & 0.0338 & 0.9828  & 205.03 / 169  \\
        QSO & 1.49 &  1.0540 & 0.0376 & 1.0105  & 458.87 / 432  \\
        \hline
        \multicolumn{6}{l}{\textbf{POST}} \\
        \hline
        LRG1 & 0.51 & 0.9761 & 0.0276 & 0.9631 & 159.45 / 121  \\
        LRG2 & 0.71 & 0.9822 & 0.0257 & 0.9792 &  125.37 / 113  \\
        LRG3 & 0.92 &  1.0013 & 0.0258 & 0.9915 &  152.23 / 144  \\
        LRG3+ELG1 & 0.93 & 0.9866 & 0.0249 & 0.9798 &  172.72 / 144  \\
        ELG2 & 1.32 & 0.9808 & 0.0407 & 0.9557 &  176.84 / 169  \\
        QSO & 1.49 & 1.0580 & 0.0282 & 1.0289 &  390.00 / 432  \\
        \hline
    \end{tabular}
    \caption{Summary of $\alpha$ constraints on DESI DR1 data. $\alpha_{\rm mean}$ denotes the transverse dilation parameter extracted from the mean of the chains. $\alpha_{\rm bf}$ is the best-fit value extracted from running the minimiser on the chains. Here $\sigma_\alpha$ is the posterior standard deviation, and $\chi^2_{\rm bf}$ is the minimizer best-fit chi-squared.}
    \label{tab:alpha_summary}
\end{table}

\subsection{Detection significance} \label{ssec:detectionsig}
\begin{figure}
    \centering
    \includegraphics[width=\linewidth]{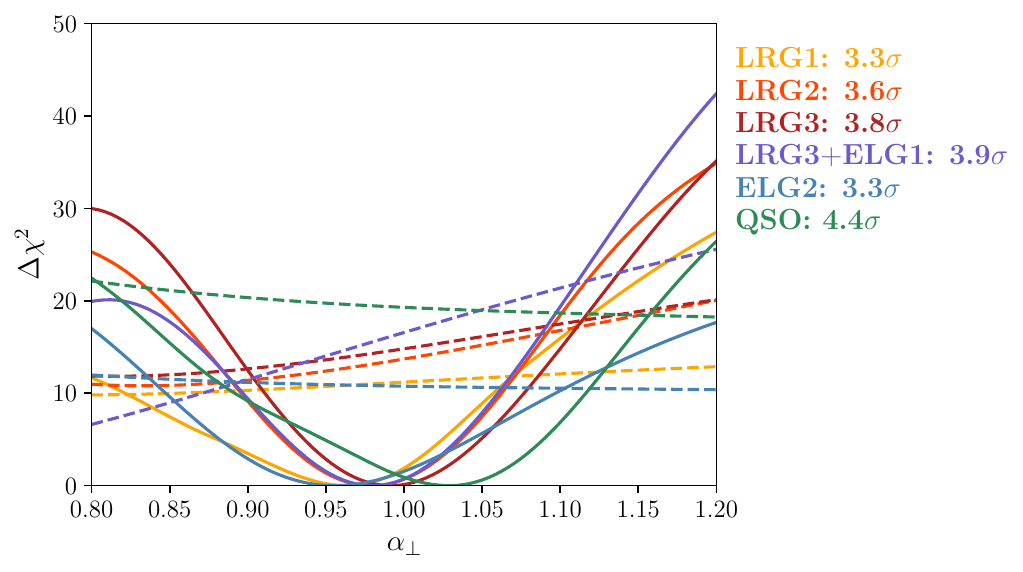}
    \caption{BAO detection significance for each DESI DR1 tracer sample. Colored curves show the $\Delta \chi^2$ as a function of $\alpha_{\perp}$ plotted relative to the minimum $\chi^2$, when considering the BAO in the fit. The dashed lines correspond to the $\Delta \chi^2$ of the no-wiggle, or no-BAO, case, subtracting the minimum of the $\chi^2$ of the BAO fit. We tabulate the detection significance in the legend, as defined by Equation \ref{eq:detection}.}
    \label{fig:deltachi2}
\end{figure}

In Figure \ref{fig:deltachi2}, analogous to Figure 11 of \cite{DESI:2024DR1}, we present the post-reconstruction $\Delta\chi^2$ profiles as a function of $\alpha_{\perp}$. The plot assesses the significance of the BAO detection by comparing the best-fit results obtained with our fiducial model, which includes the BAO peak, to those obtained with a model without the BAO feature. For each value of $\alpha_{\perp}$, we re-optimise the remaining nuisance and broadband parameters and compute the minimum $\chi^2$. 
We define $\Delta\chi^2$ as 
\begin{equation}\label{eq:deltachi2}
    \Delta\chi^2_i(\alpha_{\perp}) = \chi^2_{i}(\alpha_{\perp}) - \chi^2_{\rm wiggle,\,min},
\end{equation}
where $i = {{\rm wiggle}, {\rm no wiggle}}$. The solid curves ($i = {\rm wiggle} $) indicate how strongly the data constrains the angular BAO scale, while the dashed lines ($i = {\rm no wiggle}$) show the corresponding smooth broadband-only comparison used to quantify the BAO detection significance. The profiles are relatively broad, in comparison to the profile curves in Figure 11 of \cite{DESI:2024DR1}. That is expected, as the projection of the 3D information onto the 2D plane will broaden the BAO information and will therefore make the constraints less sharply localised than in a full 3D analysis. Following from Equation \ref{eq:deltachi2}, we quantify the detection significance as:
\begin{equation}\label{eq:detection}
    \sqrt{\Delta \chi^2_{\rm det}} = \sqrt{\Delta \chi^2_{\rm no\text{-}wiggle}(\hat{\alpha}_{\perp})},
\end{equation}
where $\hat{\alpha}_{\perp} =\arg\min_{\alpha_{\perp}}\chi^2_{\rm wiggle}$. We find a significant detection ($ > 3\sigma$) in every tracer sample, in order of effective redshift: $3.3\sigma$ for LRG1, $3.6\sigma$ for LRG2, $3.8\sigma$ for LRG3, $3.9\sigma$ for LRG3+ELG1, $3.3\sigma$ for ELG2, and $4.4\sigma$ for QSO. A comparison with the 3D BAO measurement on the same samples \cite{DESI:2024DR1} should be read with some care, since those significances are obtained from one-dimensional isotropic fits profiling $\alpha_{\rm iso}$ whereas ours profile $\alpha_{\perp}$. Given this caveat, we note that the LRG and ELG2 bins retain between $43\%$ and $53\%$ of their detection significance. The QSO bin is the exception, with an angular BAO significance comparable to its 3D BAO value. We treat the QSO detection with some caution, noting that it also sits highest relative to the mock detection among the six tracers, but nonetheless within the expectation, as shown in Appendix \ref{app:pipeval}. Unlike the 3D case, the no-wiggle profile retains a weak dependence on $\alpha_\perp$. This arises because dilation also changes the projected broadband shape, so that the no-wiggle template contains some $\alpha_\perp$-dependence even in the absence of the acoustic feature. This dependence carries no interpretable BAO-scale information. We therefore evaluate the BAO/no-BAO difference at $\hat\alpha_\perp$, the value preferred by the wiggle fit, as in Equation \ref{eq:detection}, rather than minimising the no-wiggle likelihood with respect to $\alpha_\perp$. The variation of the no-wiggle curve across the prior range quantifies the residual broadband sensitivity to dilation and should not be interpreted as a reduction in the BAO detection significance. The reason is that $\alpha_{\perp}$ enters both the wiggle and no-wiggle parts of the template in Equation \ref{eq:wtheta_model}.\footnote{A future implementation in which the split is made directly at the level of $w(\theta)$ would remove this residual broadband dependence and simplify the interpretation of the no-wiggle profile.} 

Overall, the measured significances agree well with the expectations from the mocks. For $25$ \abacus\ \altmtl realisations analysed through an identical pipeline, the mean recovered significance is $3.39\sigma$ for LRG1, $3.68\sigma$ for LRG2, $3.61\sigma$ for LRG3 and $3.10\sigma$ for ELG2, against measured values of $3.29$, $3.63$, $3.82$ and $3.29\sigma$ respectively. The QSO measurement of $4.38\sigma$ exceeds its mock mean of $3.61\sigma$ by more than any other sample, but remains within the scatter of the individual realisations. This detection can be plausibly attributed to a fluctuation of the particular realisation of the density field that we observe, and should in any case be taken with caution. The full distributions from the \abacus\ \altmtl mocks for all tracers are shown in Appendix \ref{app:pipeval}. 

\subsection{$\alpha$ values and comparison with external datasets}

\begin{figure}
    \centering
    \includegraphics[width=\linewidth]{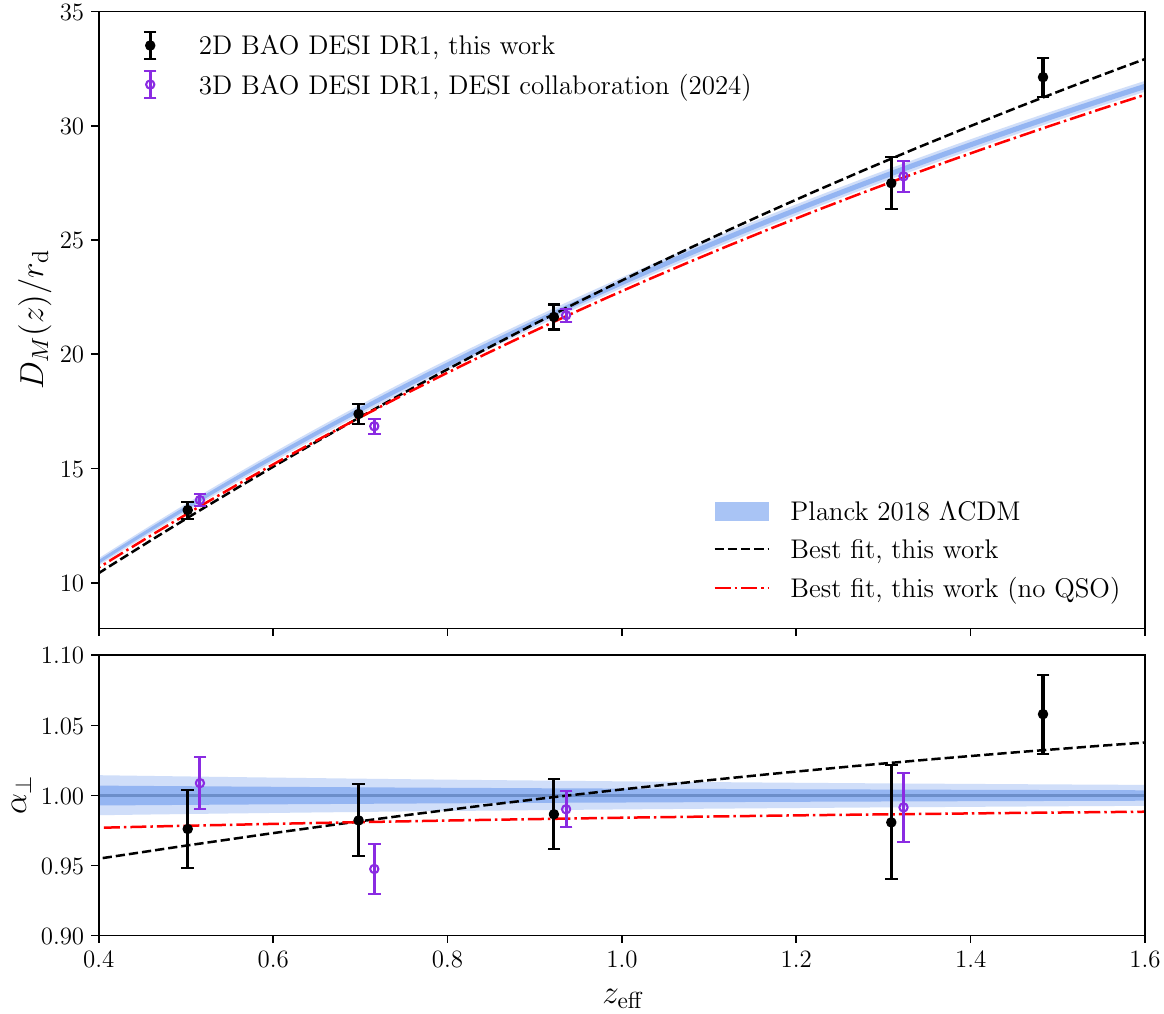}
    \caption{{\it Top panel}: Comparison between the transverse BAO distance constraints in units of the sound horizon, $D_M(z)/r_{\rm d}$, obtained from 2D and 3D BAO measurements on DESI DR1 data. Shaded bands indicate $68\%$ and $95\%$ credible intervals of the Planck 2018 $\Lambda$CDM prediction. The black dashed curve denotes the best fit $\Lambda$CDM model when fitting $\Omega_{\rm m}$ and $hr_{\rm d}$ to all the tracers considered in this work. The red dot-dashed line indicates this same fit when we remove the QSO tracer from the data vector. {\it Bottom panel}: Portrays the same information as the top panel, but relative to the fiducial Planck 2018 $\Lambda$CDM curve, so that the measurements are in terms of $\alpha_{\perp}$.}
    \label{fig:2dvs3d}
\end{figure}

In the top panel of Figure \ref{fig:2dvs3d}, we plot the transverse BAO distance constraints in units of the sound horizon as a function of the effective redshift, expressed as $D_M(z)/r_{\rm d}$, obtained from 2D BAO measurements (this work) and the corresponding DESI DR1 3D BAO results \cite{DESI:2024DR1}. We note that there is no corresponding $D_M(z)/r_{\rm d}$ measurement for the QSO sample in the official 3D DESI DR1 results. The shaded bands show the $68\%$ and $95\%$ credible intervals of the Planck 2018 $\Lambda$CDM prediction, which also defines our fiducial cosmology. The dashed and dot-dashed curves are the best-fit $\Lambda$CDM models obtained from our measurements, fitting $(\Omega_{\rm m}, h r_{\rm d})$ to all five tracer bins and to the four non-QSO bins, respectively (refer to Section \ref{sec:cosmoinf} for full results of the fit). In the bottom panel of Figure \ref{fig:2dvs3d}, we divide by the fiducial curve and visualise the results in terms of $\alpha_{\perp}$. Our five measurements reach a precision of $2.5$-$4.2\%$ on $D_M/r_{\rm d}$, with the tightest being LRG3+ELG1 at $2.5\%$ and the loosest ELG2 at $4.2\%$. The corresponding 3D DESI DR1 constraints are a factor of $1.4-2.0$ tighter at the same effective redshifts, which quantifies the information lost when the line-of-sight modes are discarded. The four non-QSO bins all lie slightly below the fiducial prediction, individually within $1\sigma$, and combine to $\alpha_\perp = 0.982 \pm 0.014$, i.e., $1.3 \,\sigma$ below unity. Taking all the tracers, the five measurements are consistent with the Planck 2018 prediction with $\chi^2 = 5.97$ for 5 degrees of freedom.

At the same time, Figure \ref{fig:2dvs3d} also illustrates the current limit of what can be achieved using the 2D information alone. The 2D constraints are derived from angular BAO features measured across individual redshift bins, and therefore remain sensitive to binning choices, covariance modelling, and in particular the statistical power available in each bin. 

The offset of the QSO point at $z_{\rm eff} = 1.49$ should be interpreted with this in mind. It sits $2.1 \sigma$ above the Planck 2018 prediction and $2.4\sigma$ above the combination of the four lower-redshift bins, and because it is a high-redshift anchor for the curve, it pulls the inferred $\Omega_{\rm m}$ down substantially. Removing it shifts the mean posterior value from $\Omega_{\rm m} = 0.198^{+0.045}_{-0.068}$ to $0.212^{+0.090}_{-0.18}$. Given the sparsity of the DR1 quasar sample in this shell, the offset is consistent with being a statistical fluctuation or a residual systematic in the angular BAO extraction rather than a genuine departure from Planck 2018 $\Lambda$CDM. DESI DR2 has already increased by $1.8$ the effective volume of the QSO sample over this redshift range \cite{DESI:2025DR2}, so this point can be scrutinised directly rather than left as a caveat. We note that the QSO measurement from DESI DR2 \cite{DESI:2025DR2} gives $D_M/r_{\rm d} = 30.51 \pm 0.76$, which is roughly $\alpha_\perp \simeq 1.008 \pm 0.025$ with respect to our fiducial cosmology, and is consistent with the Planck 2018 prediction at the $0.3\sigma$ level. As DR2 contains the DR1 quasars together with a substantially larger sample, the DR2 measurement gives an indication of how our DR1 QSO data point could evolve with more data.

The comparison with the DESI DR1 3D analysis in Figure \ref{fig:2dvs3d} provides a direct same-sample test of whether the angular and 3D methodologies recover consistent transverse BAO distances. Using supernovae as a redshift interpolator and assuming only the validity of the Etherington relation, \cite{Favale:2024sdq} quantified an apparent discrepancy between previous 2D and DESI DR1 3D BAO measurements. Here, by applying the angular methodology to the same DESI DR1 catalogues from which the 3D constraints of \cite{DESI:2024DR1} were extracted, we remove differences in the underlying galaxy sample and isolate differences associated with the analysis methodology. We find that the transverse distances recovered by the two approaches are consistent within the statistical precision of the present measurements in every overlapping redshift bin. We therefore find no evidence for a detectable method-dependent shift in the recovered transverse BAO scale at DESI DR1 precision.

The numerical differences between the two estimates are -1.2, +1.4, -0.2 and -0.2 times the uncertainty of the angular measurement for LRG1, LRG2, LRG3+ELG1 and ELG2, respectively. Because the two estimators are derived from the same underlying density field, their statistical errors are correlated, and these numbers should not be interpreted as formal tension significances. The agreement is most direct at $z_{\rm eff}=0.51$, where the two analyses share the same effective redshift and can be compared without interpolation. We measure $\alpha_\perp(z_{\rm eff}=0.51)=0.976\pm0.028$, or $D_M(z_{\rm eff}=0.51)/r_{\rm d}=13.18\pm0.37$, compared to $\alpha_\perp(z_{\rm eff}=0.51)=1.009\pm0.019$ from the 3D analysis.

\begin{figure}
    \centering
    \includegraphics[width=\linewidth]{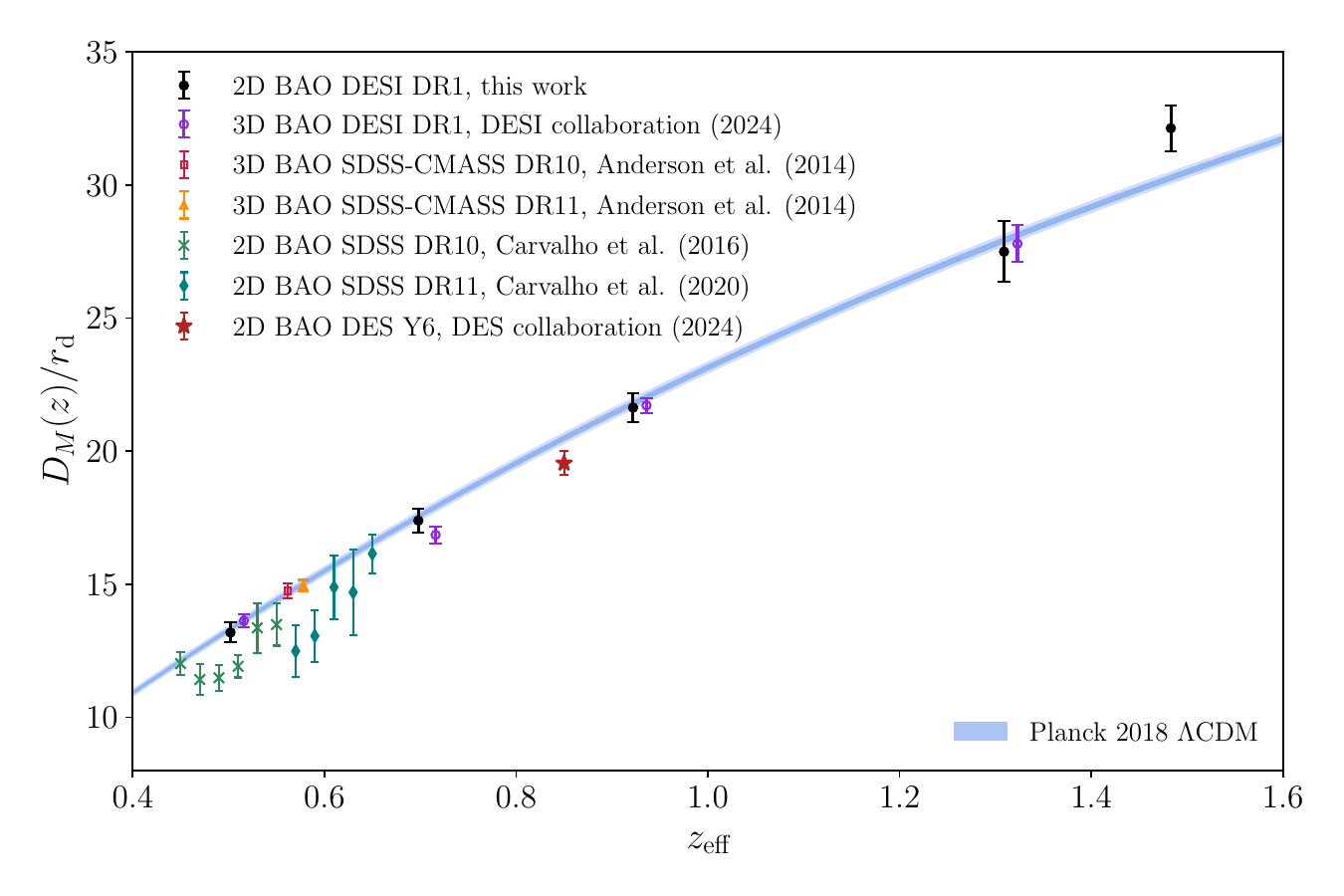}
    \caption{Comparison between the transverse BAO distance constraints in units of the sound horizon, $D_M(z)/r_{\rm d}$, obtained from this work, the DESI 3D BAO measurement as well as external 2D BAO measurements. The black solid points denote the measurements from this work, from the following tracers: LRG1, LRG2, LRG3$+$ELG1, ELG2 and QSO. The transparent purple points denote the DESI DR1 3D BAO measurements, using only the transverse component \cite{DESI:2024DR1}. The open red box marker and orange triangular marker denote the 3D BAO measurements from SDSS-III BOSS CMASS DR10 and DR11 at $z_{\rm eff}=0.57$ \cite{BOSS:2013rlg}, obtained from their post-reconstruction consensus $\alpha_\perp$ and converted to $D_M/r_{\rm d}$ with their fiducial cosmology. These two points are slightly displaced horizontally for visibility. The green cross data points are the 2D BAO measurements from SDSS DR10 \cite{Carvalho:2015ica}, the cyan diamond data points are from SDSS DR11 \cite{Carvalho:2017tuu}, and the dark red star denotes the 2D BAO measurement from the DES collaboration \cite{DES:20242dbao}. The shaded band indicates the $68\%$ and $95\%$ credible intervals of the fiducial cosmology used in this work, Planck 2018 $\Lambda$CDM.}
    \label{fig:dmrd_external}
\end{figure}

In Figure \ref{fig:dmrd_external}, we plot the comparison between our measurements and previous angular BAO works alongside the Planck 2018 $\Lambda$CDM band as well as comparisons with other 3D BAO works \cite{BOSS:2013rlg, DESI:2024DR1}. Our measurements represent a substantial improvement over previous angular BAO constraints at comparable redshifts, mainly because we do a joint fit over the individual redshift bins making up the tracer sample. The SDSS DR10 and DR11 angular measurements of \cite{Carvalho:2015ica} and \cite{Carvalho:2017tuu}, respectively, which cover $0.45 \le z \le 0.65$ in shells of $\Delta z = 0.02$, reach a fractional precision on $D_M/r_{\rm d}$ of $3.5$-$7.0\%$ and $4.5$-$11.0\%$, respectively, compared with our aggregate precision of $2.8\%$ for our LRG1 measurement at $z_{\rm eff} = 0.51$. At the LRG1 effective redshift, we find consistency with the Planck 2018 $\Lambda$CDM prediction within $1 \sigma$, whereas \cite{Carvalho:2015ica} and \cite{Carvalho:2017tuu} are discrepant within the range $2 - 4 \sigma$ for each point. We note that the difference in our results may be attributed to the fact that they used a phenomenological model to fit their measurements, whereas we used a template-fitting method. The most precise external measurement is the DES Y6 angular BAO constraint at $z_{\rm eff} = 0.851$, $\alpha_\perp (z_{\rm eff}=0.851) = 0.957 \pm 0.020$ \cite{DES:20242dbao}, obtained by fitting a single dilation parameter jointly across six tomographic bins ($0.6 < z < 1.2$) and defined with respect to the same Planck 2018 $\Lambda$CDM reference as our own. Additionally, \cite{DES:20242dbao} uses the same template-fitting pipeline, which makes their measurements directly comparable to our $\alpha_\perp$. It agrees with our nearest measurement, LRG3+ELG1, $\alpha_\perp (z_{
\rm eff}= 0.93) = 0.9866 \pm 0.0249$, at $0.9\sigma$.  

\section{Cosmological interpretation}\label{sec:cosmoinf}

We can now translate the measurements of $\alpha_{\perp}(z)$ in each redshift bin into constraints on the background expansion history. For the cosmological interpretation, we approximate the marginalized $\alpha_{\perp}$ constraints as Gaussian. We have checked the validity of this compression by comparing the 68$\%$ central and minimum credible intervals of the marginalized posteriors. For a Gaussian posterior, the two intervals coincide, while any skewness shifts the minimum interval towards the peak. The two intervals agree to the third decimal place for all tracer samples, including QSO, which corresponds to at most a few percent of the statistical uncertainty on $\alpha_\perp$. The marginalized posteriors are therefore symmetric to a good approximation within the 68$\%$ region, and the mean and standard deviation provide an adequate summary of the constraints.

We emphasise that the angular BAO analysis delivers a single geometric quantity per bin, $D_M(z)/r_{\rm d}$, and that the interpretation in what follows is designed to make explicit exactly how much cosmological information that quantity carries, both on its own and relative to the full anisotropic (3D) BAO analysis of the same data. 
We use the cosmological inference code ${\tt cobaya}$ \cite{Torrado:2020dgo} to perform the MCMC, and to compute the theoretical cosmology calculations, we rely on the Boltzmann code {\tt CAMB} \cite{Lewis:1999bs}. 

From the BAO-fitting code, we obtain the transverse dilation parameter $\alpha_{\perp}$, defined with respect to the fiducial cosmology,
\begin{equation}
    \alpha_\perp(z_i) = \frac{D_M(z_i)/r_{\rm d}}{\left[D_M(z_i)/r_{\rm d}\right]^{\mathrm{fid}}}\, ,
\end{equation}
which we invert to build the data vector ${\bf d}$ as well as the corresponding errors obtained from the BAO fitting MCMC, with the fiducial ratio a fixed constant (determined by the fiducial cosmology),
\begin{align}
\left(\frac{D_M}{r_{\rm d}}\right)^{\mathrm{obs}}_{i}
    &= \alpha_{\perp,i}\left(\frac{D_M}{r_{\rm d}}\right)^{\mathrm{fid}}_{i}\, ,
    &
    \sigma_i &= \sigma_{\alpha_\perp,i}\left(\frac{D_M}{r_{\rm d}}\right)^{\mathrm{fid}}_{i}\,.
\end{align}
The index $i$ here labels the different tracer samples that we study in this work. The data vector used throughout this section is, therefore, 
\begin{equation}
    {\bf d} = \Big\{ (D_M/r_{\rm d})^{\mathrm{obs}}_i \Big\}\, ,\quad
    i \in \{\mathrm{LRG1},\,\mathrm{LRG2},\,\mathrm{LRG3{+}ELG1},\,\mathrm{ELG2},\,\mathrm{QSO}\}\,,
\end{equation}
with effective redshifts $z_i = 0.51,\,0.71,\,0.93,\,1.32,\,1.49$.
We consider two sets of data vectors, the first being all the tracers considered in this work: LRG1, LRG2, LRG3$+$ELG1, ELG2, and QSO. This will showcase the total cosmological information that the angular BAO analysis can quantify. Secondly, we consider the non-QSO case, whereby we omit the QSO tracer from the fit. This is for the purposes of direct comparison to the DESI DR1 3D $D_M/r_{\rm d}$ BAO results, as the QSO tracer does not have the $D_M/r_{\rm d}$ equivalent in the 3D BAO case of DR1. We also do not compare with the DESI DR1 3D Ly$\alpha$ forest measurement of $D_M/r_{\rm d}$ and we therefore omit it in all relevant fits.

To construct the theory vector $\boldsymbol{t}$, we assume a spatially flat $\Lambda$CDM background ($w=-1$, $\Omega_k = 0$) with a fixed neutrino mass sum $\sum m_\nu = 0.06\,\mathrm{eV}$. The transverse BAO measurements then constrain the comoving angular diameter distance relative to the sound horizon at the drag epoch,
\begin{equation}\label{eq:dmrD_Model}
    \left(\frac{D_M}{r_{\rm d}}\right)(z;\boldsymbol{\Theta})
    = \frac{c}{H_0\, r_{\rm d}}\int_0^z \frac{\mathrm{d}z'}{E(z';\Omega_{\rm m})}\, ,
\end{equation}
where $E(z) = \sqrt{\Omega_{\rm m}(1+z)^3 + \Omega_r(1+z)^4 + 1 - \Omega_{\rm m} - \Omega_r}$, $\boldsymbol{\Theta}$ is the free cosmological parameter vector and $r_{\rm d} = \int_{z_d}^{\infty} c_s(z)/H(z)\,\mathrm{d}z$ is computed self-consistently from $\boldsymbol{\Theta}$. The radiation density $\Omega_{\rm r}$ is not a free parameter. It is fixed by the CMB temperature, $T_{\rm CMB}=2.7255\,{\rm K}$, and the effective number of relativistic species, $N_{\rm eff}=3.044$, and computed consistently by {\tt CAMB}. We note that it is only relevant for the BBN-calibrated parameterisation, which we specify later on. Written in this way, Equation \ref{eq:dmrD_Model} makes the structure of the problem explicit. The shape of the distance-redshift relation is controlled by $\Omega_{\rm m}$ alone, while $H_0$ and $r_{\rm d}$ enter only through their product. We exploit this by running two distinct parameterisations which differ in the assumptions they make about the sound horizon and which we report separately.

In what follows, we adopt a Gaussian likelihood with a diagonal covariance across the tracer bins,
\begin{equation}
    -2\ln\mathcal{L}(\boldsymbol{\Theta})
    = \big[{\bf d}-{\bf t}(\boldsymbol{\Theta})\big]^{\!\top}
      {\sf C}^{-1}
      \big[{\bf d}-{\bf t}(\boldsymbol{\Theta})\big],
    \qquad
    {\sf C} = \mathrm{diag}\big({\sf C}_{\rm LRG1},\dots,{\sf C}_{\rm tracer}\big).
\end{equation} We tabulate the priors and fixed values in Table \ref{tab:priors}. We now consider two choices of $\boldsymbol{\Theta}$ and therefore the physics that is assumed to compute $\boldsymbol{t}(\boldsymbol{\Theta})$.
 
\paragraph{Geometric (uncalibrated) parameterisation.} Writing $H_0 r_{\rm d} = 100\,(h\,r_{\rm d})\,\mathrm{km\,s^{-1}}$ with $h\,r_{\rm d}$ in Mpc,  we sample $\boldsymbol{\Theta} = (\Omega_{\rm m},\, h\,r_{\rm d})$ directly, with $H_0$ held at a fixed dummy value that cancels identically from Equation \ref{eq:dmrD_Model}. This is the honest parameterisation for a standard-ruler measurement as it assumes only that the BAO feature is a ruler of fixed comoving length, makes no statement about the value of that length, and therefore imposes no early-Universe physics. The results in this parameterisation are the primary cosmological product of this work.

\paragraph{BBN-calibrated parameterisation.} To break the $H_0$-$r_{\rm d}$ degeneracy and quote a measurement of $H_0$ only, we adopt the standard BAO$+$BBN construction of \cite{DESI:2024cosmo}. We sample $(\Omega_{\rm m},\, H_0,\, \omega_{\rm b})$, place a Gaussian prior on $\omega_{\rm b}$ from primordial light-element abundances \cite{Schoneberg:2024ifp}, reconstruct $\omega_\mathrm{cdm} = \Omega_{\rm m}h^2 - \omega_{\rm b} - \omega_\nu$, and let {\tt CAMB} integrate $r_{\rm d}(\omega_{\rm b}, \omega_\mathrm{cdm})$ at every step of the chain. The effective free parameters are $\boldsymbol{\Theta} = (\Omega_{\rm m},\, H_0)$. This is a strictly stronger set of assumptions than the geometric case, presuming standard BBN, $N_\mathrm{eff} = 3.044$ and standard pre-recombination physics, and we report this result for the sake of completeness.

\begin{figure}
    \centering
    \includegraphics[width=\linewidth]{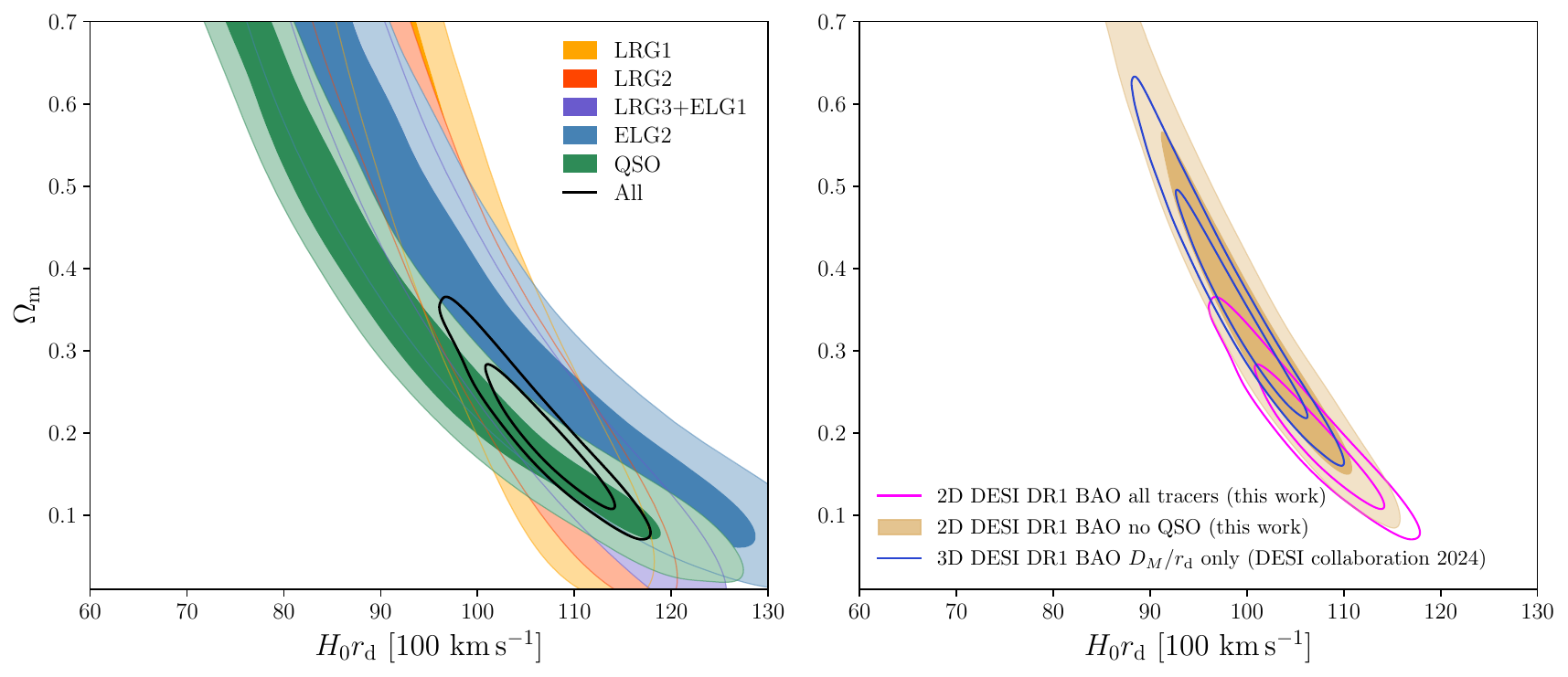}
    \caption{{\it Left panel}: $68\%$ and $95\%$ credible interval contours in the parameter space ($\Omega_{\rm m}, H_0r_{\rm d})$ from performing flat $\Lambda$CDM fits to the angular BAO measurements for each tracer type in this work. The black transparent contour denotes the constraining power when we take the data vector to contain all the transverse BAO measurements of all the tracers. The change in degeneracy direction is as expected due to the different redshifts of the sample. {\it Right panel}: Comparison between the $68\%$ and $95\%$ credible interval contours when considering all tracers in this work (magenta), when we omit the QSO data point from the fit (orange), and the transverse component of the 3D counterpart from DESI official (blue). The latter does not report a transverse measurement for the QSO sample and we have omitted the Ly$\alpha$ measurement. The most direct comparison is between the blue and orange contours, which are in excellent agreement. }
    \label{fig:omhrd}
\end{figure}

In Figure \ref{fig:omhrd}, we show the constraints in the $(H_0 r_{\rm d},\,\Omega_{\rm m})$ plane. On the left panel, we show the constraints tracer by tracer and we see that it is highly degenerate because a single value of $D_M(z)/r_{\rm d}$ fixes a continuous family of $(H_0 r_{\rm d},\,\Omega_{\rm m})$ pairs. The orientation of that band is set by the sensitivity of the distance integral to $\Omega_{\rm m}$, which vanishes as $z\to0$ and grows steadily with redshift, so the bands rotate from nearly vertical for the low-redshift LRG samples towards the horizontal for ELG2 and QSO. The combined contour for all tracers shown in black is produced by the intersection of differently tilted bands, which means that the highest-redshift bins carry the most leverage on $\Omega_{\rm m}$. This is visible in the right panel, where removing the QSO sample, whose $\alpha_\perp$ sits farthest above unity, shifts $\Omega_{\rm m}$ appreciably while the shift in $H_0 r_{\rm d}$ is milder, as expected for a parameter set by the overall amplitude rather than the shape of the distance-redshift relation. Additionally, the right panel isolates how much cosmological information $\alpha_\perp$ carries on its own. The blue contour uses the published DESI DR1 anisotropic measurements with the $D_H/r_{\rm d}$ rows discarded, fitted on the same bins with the same likelihood and priors, so it differs from our result only in how $\alpha_\perp$ was measured. The two contours share the same degeneracy direction and are statistically consistent, with our constraint being the broader of the two. This is expected, since the angular analysis measures $\alpha_\perp$ from the projected clustering in each redshift shell and recovers it with larger per-bin uncertainties than the anisotropic fit, which uses the full 3D clustering. The value of the comparison lies not in matching the precision of the 3D analysis but in isolating the transverse information, and in demonstrating that the two different routes to $\alpha_\perp$ agree on the cosmology they imply. We tabulate the results in Table \ref{tab:cosmo_results} with the different combinations of tracers.

\begin{figure}
    \centering
    \includegraphics[width=0.6\linewidth]{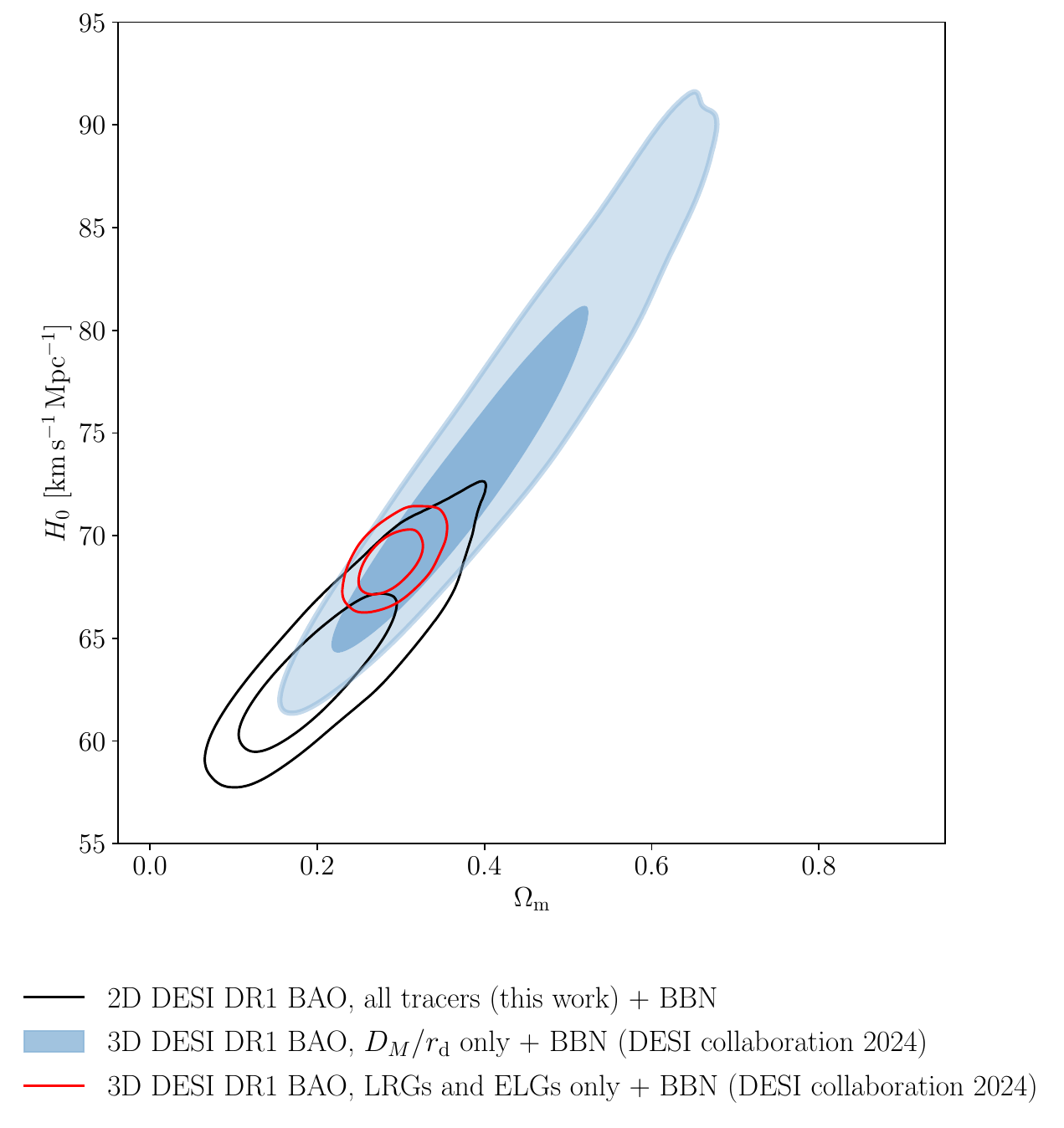}
    \caption{$68\%$ and $95\%$ credible intervals in the parameter space ($\Omega_{\rm m}, H_0$) when performing fits using a flat $\Lambda$CDM model, with the additional prior on $\omega_{\rm b}$ determined by BBN. We visualise the contours for this work when all tracers are considered (black), the transverse component of the 3D counterpart from DESI DR1 official results (blue), and the full anisotropic 3D BAO measurement from DESI DR1 only for LRG1, LRG2, LRG3+ELG1 and ELG2 (red). All contours are in agreement within $1\sigma$.}
    \label{fig:bbn}
\end{figure}

Figure \ref{fig:bbn} shows the effect of calibrating the sound horizon with the BBN prior on $\omega_{\rm b}$, which converts the constraint on the product $H_0 r_{\rm d}$ into one on $H_0$ itself. Adding BBN does not, however, break the degeneracy completely. Both our measurement in black and the DESI transverse-only vector in blue remain extended along the $\Omega_{\rm m}$--$H_0$ direction, with strongly asymmetric posteriors and uncertainties several times larger than those obtained from the full anisotropic analysis on the same bins, shown in red, which gives $\Omega_{\rm m} = 0.290 \pm 0.026$ and $H_0 = 68.8 \pm 1.1\ \mathrm{km\,s^{-1}\,Mpc^{-1}}$. This behaviour is reproduced identically whether the transverse-only vector is ours or the 3D vector from the official DESI paper. The transverse dilation parameter constrains a single distance at each redshift, and calibrating the ruler with BBN fixes the amplitude of that distance without adding any independent handle on its shape. Breaking the degeneracy requires the radial information, since $\alpha_\parallel(z)$ constrains $H(z) r_{\rm d}$ along a direction that is not parallel to the transverse one, and the two together constrain the contour far more efficiently than either does alone. All three contours overlap, and the cosmologies inferred from the transverse-only and full anisotropic analyses are mutually consistent.

\begin{table}
    \centering
    \begin{tabular}{lll}
        \hline
        Parameter & Geometric & BBN-calibrated \\
        \hline
        $\Omega_{\rm m}$          & $\mathcal{U}(0.01,\,0.99)$ & $\mathcal{U}(0.01,\,0.99)$ \\
        $h\,r_{\rm d}$ [Mpc]      & $\mathcal{U}(10,\,1000)$   & derived quantity \\
        $H_0$ [$\mathrm{km\,s^{-1}Mpc^{-1}}$] & fixed (cancels)   & $\mathcal{U}(20,\,100)$ \\
        $\omega_{\rm b}$          &  $0.02207$           & $\mathcal{N}(0.02218,\,0.00055^2)$ \\
        $r_{\rm d}$               & -  & derived from \texttt{CAMB} \\
        $\sum m_\nu$ [eV]            & $0.06$             & $0.06$ \\
        $w$, $w_a$, $\Omega_k$       & $-1$, $0$, $0$      & $-1$, $0$, $0$  \\
        \hline
    \end{tabular}
    \caption{Parameters and priors. $\mathcal{U}(a,b)$ denotes a uniform prior on $[a,b]$ and $\mathcal{N}(\mu,\sigma^2)$ a Gaussian prior. The two calibrations are run separately. Parameters not sampled in a given calibration are either fixed (geometric $H_0$, which cancels identically) or derived from the theory code ($r_{\rm d}$ in the BBN calibration).}
    \label{tab:priors}
\end{table}

\begin{table}
    \centering
    \begin{tabular}{lcccc}
        \hline
        Sample & $\Omega_{\rm m}$ & $H_0 r_{\rm d}$ [100 km\,s$^{-1}$] & $H_0$ [km\,s$^{-1}$Mpc$^{-1}$]  \\
        \hline
        \multicolumn{4}{l}{\bf Geometric}\\
        All tracers  & $0.198^{+0.045}_{-0.068}$ & $107.0^{+4.7}_{-4.3}$ & ---  \\
        LRGs, ELGs & $0.360^{+0.090}_{-0.183}$ & $99.8^{+7.4}_{-6.2}$ & ---\\
        \hline
        \multicolumn{4}{l}{\bf BBN-calibrated}\\
        All tracers  & $0.212^{+0.047}_{-0.079}$ & ---  & $63.9^{+1.9}_{-3.3}$  \\
        LRGs, ELGs & $0.407^{+0.118}_{-0.182}$ & ---  & $75.8^{+5.7}_{-11.8}$  \\
        \hline
    \end{tabular}
    \caption{Marginalised constraints on the background expansion from the angular BAO measurements presented in this work. All parameters that have been tabulated are varied parameters. Uncertainties are $68\%$ credible intervals. \emph{All tracers} denotes the full five-bin data vector, \emph{LRGs, ELGs} denotes the following tracers in the data vector: LRG1, LRG2, LRG3+ELG1, ELG2. }
    \label{tab:cosmo_results}
\end{table}

\section{Conclusions}\label{sec:conc}
 
This work has presented a tomographic angular BAO analysis of the DESI DR1 large-scale structure catalogues, measuring the two-point angular correlation function in thin redshift shells across five tracer samples spanning $0.4 < z < 2.1$. By construction, this approach requires no fiducial cosmology to convert redshifts into comoving distances along the line of sight, and it therefore constrains the transverse dilation parameter $\alpha_\perp(z)$ through a route that is largely independent of the coordinate mapping underlying the standard 3D analysis. We validated the pipeline extensively on \abacussummit mocks, characterising the expected detection significance and the recovery of the fiducial $\alpha_\perp(z)$ value, before applying it to the DESI DR1 data.

We obtain a significant BAO detection in every tracer bin, exceeding $3\sigma$ throughout, and measure $\alpha_\perp(z)$ to a precision of $2.5$--$4\%$ per bin. Compared directly with the DESI DR1 3D transverse constraints obtained from the same tracer samples, the angular measurements are consistent within the statistical precision of DR1, with no evidence for a detectable method-dependent offset in the recovered transverse BAO scale. Since both analyses use the same tracer samples and constrain the same transverse distance, this comparison largely removes sample-selection differences and directly tests whether the distinct analysis methodologies lead to different recovered transverse BAO scales. Within the statistical precision of DESI DR1, we detect no such method-dependent offset. 

The absence of a detectable same-sample offset also bears on interpretations of previously reported 2D–3D BAO discrepancies in terms of non-standard cosmology \cite{Ruchika:2024lgi, Bernui:2023byc, Pantos:2026rpe, Dwivedi:2024okk}. A generic modification of the background expansion shifts the physical transverse distance and would therefore affect both methods in the same way. Our comparison does not constrain such models beyond the usual BAO distance information. It does, however, constrain explanations in which the reported discrepancy arises because the angular and 3D analyses respond differently to non-standard geometry or to assumptions entering the 3D distance mapping. At the precision of DESI DR1, we find no evidence for such an effect. Interpreting the angular measurements cosmologically, a flat $\Lambda$CDM fit to $\alpha_\perp$ alone yields $\Omega_{\rm m} = 0.198^{+0.045}_{-0.068}$, a result that sits $1.6\sigma$ below the DESI 3D value. This pull is driven almost entirely by the highest-redshift bin, the quasar sample, where $\alpha_\perp(z_{\rm eff} = 1.49) = 1.058 \pm 0.028$ lies $2\sigma$ above the fiducial cosmology. Restricting the fit to the four lower-redshift bins for which DESI reports a transverse-only measurement \cite{DESI:2024DR1}, our inferred $\Omega_{\rm m}$ and $H_0 r_{\rm d}$ agree with the corresponding 3D fit to better than $0.1\sigma$, indicating that the offset is confined to the quasar bin rather than reflecting a systematic shared across the sample. Given the limited statistics of the DR1 quasar catalogue in this shell, we consider a statistical fluctuation, or a residual systematic specific to this bin, a more plausible explanation than a genuine departure from $\Lambda$CDM. With a BBN prior breaking the $H_0-r_{\rm d}$ degeneracy, the full five-bin sample gives $H_0=63.9^{+1.9}_{-3.3}\,{\rm km\,s^{-1}\,Mpc^{-1}}$. Unlike the geometric $H_0r_{\rm d}$ constraint, this inference is calibrated and assumes standard BBN and pre-recombination physics.

Ultimately, this work demonstrates that a tomographic angular BAO analysis at the precision needed to test the 3D pipeline is achievable with present DESI data, a capability earlier surveys lacked. The pipeline developed here is directly applicable to forthcoming DESI data releases with a small modification for how the no-wiggle template is implemented at the BAO-fitting stage. DESI DR2 will increase the effective volume of the quasar sample by $1.8$ \cite{DESI:2025DR2} over the redshift range considered in this work, which will determine whether the offset in the $z_{\rm eff} = 1.49$ bin persists as a genuine feature or resolves as expected under improved statistics, while the corresponding growth in sample size across all tracers will tighten the angular BAO constraints throughout. More broadly, as photometric surveys such as LSST and Euclid reach maturity, the validation presented here lends direct support to the use of angular BAO measurements as a stand-alone cosmological probe in surveys without access to precise spectroscopic redshifts.

\acknowledgments

AP acknowledges support of the ``la Caixa'' Foundation fellowship (ID 100010434). The fellowship code is ``LCF/BQ/DI24/12070011''. AP, HGM and LV  acknowledge support of Proyectos de Generación de Conocimiento (PID2022-141125NB-I00) of the Spanish Ministry of Science and Innovation and the ``Center of Excellence Maria de Maeztu 2025-2029'' award to the ICCUB funded by grant CEX2024-001451-M from AEI/10.13039/501100011033

\appendix

\section{Validation of pipeline on mocks}\label{app:pipeval}

All the analysis choices described in Section \ref{ssec:baofitting} were validated and fixed using the 25 \abacus\ \altmtl mock catalogues before the pipeline was applied to the DESI DR1 measurements. This includes the number of broadband terms, the angular fitting ranges and the redshift and angular binning. None of these choices were revisited after the data were analysed. In the following appendix, we present the mock-based tests that set the fiducial choices in the main analysis. For simplicity and ease of reading, we use $\alpha$ instead of $\alpha_{\perp}$ notation-wise in the following sections.

\subsection{Choice of broadband terms}\label{app:bbterms}

The number of broadband terms in Equation \ref{eq:wtheta_model} was fixed using the mock tests presented here, before the pipeline was applied to the DESI measurements. The template $w_{\rm temp}^i(\theta)$ is computed in the fiducial cosmology, and its shape away from the BAO peak need not match that of the measured $w(\theta)$. The broadband terms are included to account for this difference in shape. This matters for the BAO measurement because $\alpha$ rescales the angular coordinate of the whole template, and therefore changes not only the position of the peak but also the slope of the template across the fitting window. If the broadband terms cannot account for a difference in slope between the template and the data, the fit can partly compensate by shifting $\alpha$, which biases the recovered BAO scale. Each additional broadband term reduces this risk but adds a free parameter that is partially degenerate with $A^i$ and $\alpha$. With too many terms, the broadband contribution can also follow noise fluctuations in the measurement, including those close to the peak, which broadens the posterior on $\alpha$ and makes the fit less stable between realisations. We assess this trade-off on mocks, for which the expected value of $\alpha$ is known.

We compare two configurations of Equation \ref{eq:wtheta_model}. In the first, which we refer to as 1BB, the broadband contribution in each redshift bin is a single constant $a_0^i$. In the second, which we refer to as 2BB, we keep both the constant and the $\theta^{-1}$ term, so that the broadband contribution is $a_0^i + a_1^i\,\theta^{-1}$. This corresponds to truncating the sum in Equation \ref{eq:wtheta_model} at $j=1$. For each configuration, we run the joint fit described in Section \ref{ssec:baofitting} on 25 \abacus mocks for every tracer, both before and after reconstruction. The template is computed in the same Planck 2018 cosmology as the mocks, so an unbiased pipeline should return $\langle\alpha\rangle = 1$.

\begin{figure}
    \centering
    \includegraphics[width=\linewidth]{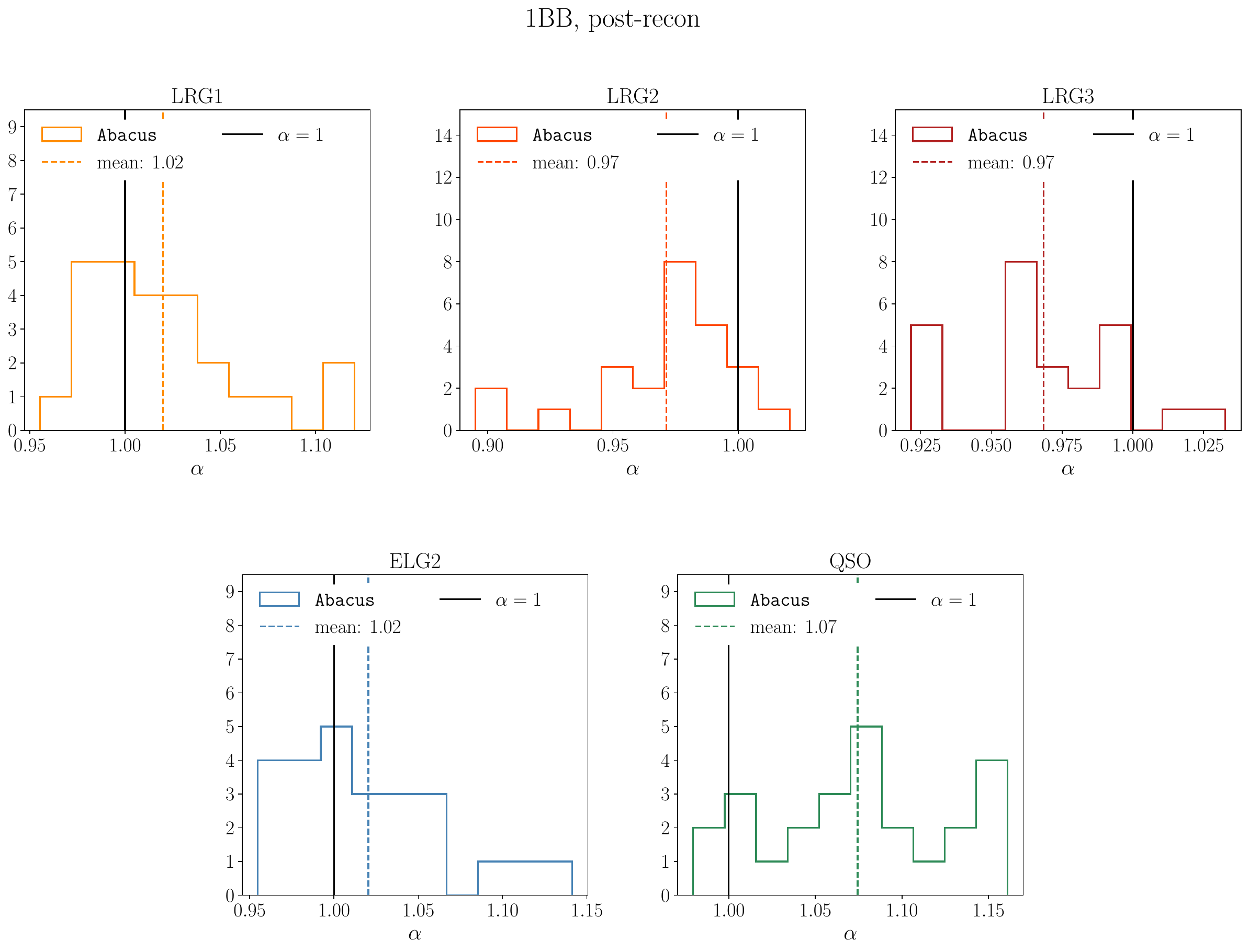}
    \caption{Distribution of the best-fit $\alpha$ across 25 \texttt{Abacus} mocks for each tracer, using the 1BB broadband model on the post-reconstruction measurements. The solid black line marks the expected value $\alpha = 1$ and the dashed line marks the mean over the mocks.}
    \label{fig:alpha_1bb_post}
\end{figure}

\begin{figure}
    \centering
    \includegraphics[width=\linewidth]{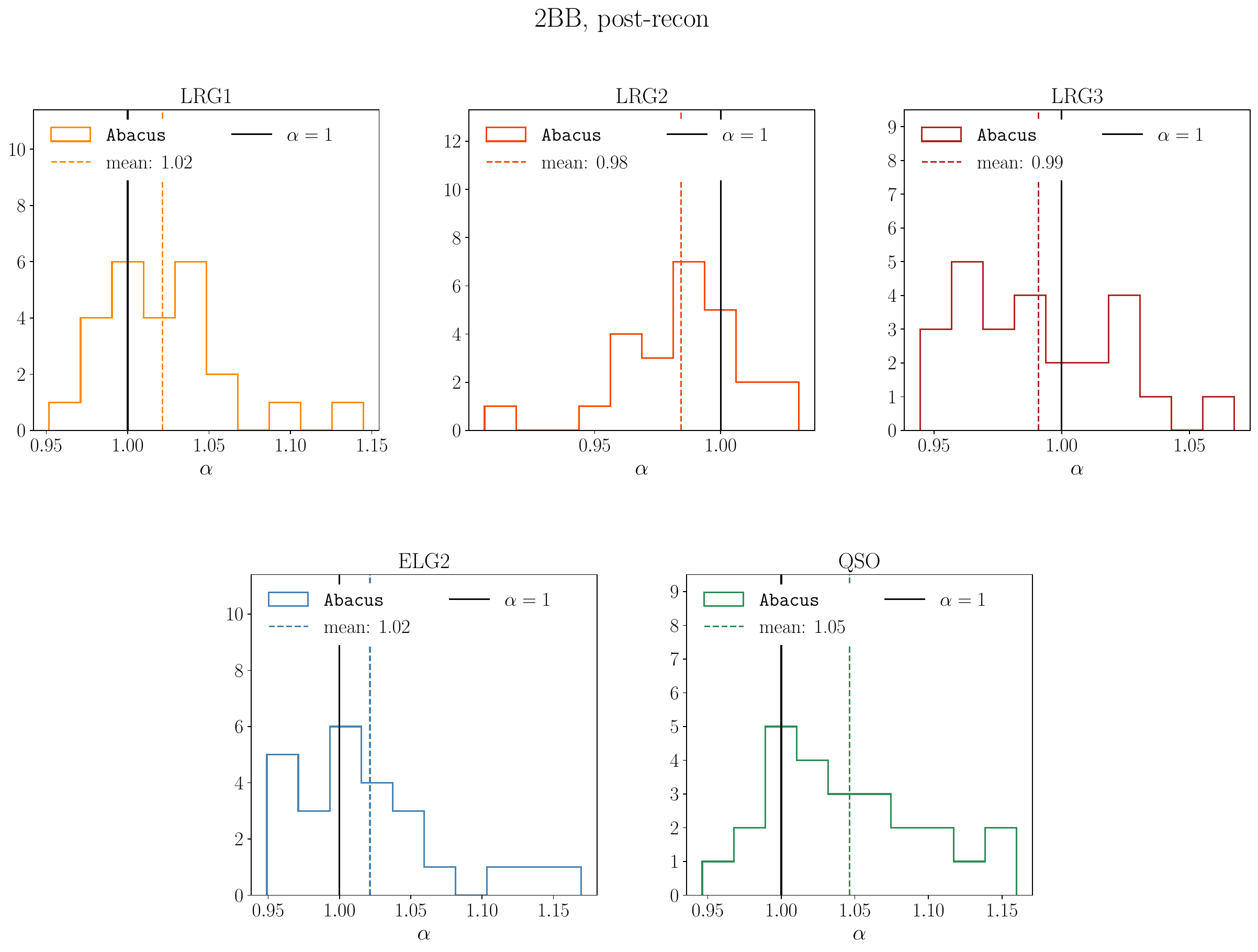}
    \caption{Same as Figure~\ref{fig:alpha_1bb_post} but for the 2BB broadband model adopted in the fiducial analysis.}
    \label{fig:alpha_2bb_post}
\end{figure}

Figures~\ref{fig:alpha_1bb_post} and \ref{fig:alpha_2bb_post} show the distribution of the best-fit $\alpha$ across the 25 mocks for each tracer in the 1BB and 2BB configurations, respectively, using the post-reconstruction measurements as an example. The full set of results, including the pre-reconstruction case, is given in Table \ref{tab:1bb2bbtable}. For each configuration, we report the mean $\langle\alpha\rangle$ and its offset from unity, the mean posterior uncertainty $\langle\sigma_{\alpha}\rangle$ and the standard deviation of $\alpha_\perp$ across the mocks. After reconstruction, the 2BB model recovers $\alpha$ more accurately than the 1BB model for LRG2, LRG3 and QSO. The offset from unity decreases from $-0.029$ to $-0.016$ for LRG2, from $-0.032$ to $-0.009$ for LRG3 and from $+0.074$ to $+0.046$ for QSO. For LRG3 and QSO this improvement corresponds to roughly $0.8\,\langle\sigma_\alpha\rangle$. For LRG1 and ELG2 the two configurations give essentially the same mean. Before reconstruction, the largest difference is again for QSO, where the offset is reduced from $+0.078$ to $+0.040$ when the second broadband term is included. For LRG2, LRG3 and ELG2 the pre-reconstruction offsets are slightly larger with 2BB, but the increase in $|\langle\alpha\rangle - 1|$ is at most about $0.15\,\langle\sigma_\alpha\rangle$ and is much smaller than the improvements seen for the other samples.

The additional freedom of the 2BB model comes at a modest cost in precision. The mean posterior uncertainty increases by at most about 20 per cent before reconstruction and about 13 per cent after reconstruction, and for ELG2 and QSO post-reconstruction it is slightly smaller with 2BB. This indicates that the $\theta^{-1}$ term does not absorb a significant part of the BAO signal. Taken together, these results show that a single constant is not flexible enough to describe the broadband shape of $w(\theta)$ for all tracers, and that adding the $\theta^{-1}$ term removes most of the resulting bias while leaving the constraining power largely intact. We therefore adopt the 2BB configuration in the fiducial analysis.

\begin{table}
    \centering
    \begin{tabular}{llcccc}
        \toprule
        BB & Tracer & $\langle\alpha\rangle$ & $\langle\alpha\rangle - 1$ & $\langle\sigma_{\alpha}\rangle$ & $\sigma(\alpha)$ \\
        \midrule
        \multicolumn{6}{l}{\bf PRE} \\
        \midrule
        1BB & LRG1 & 1.0239 & $+0.0239$ & 0.0383 & 0.0448 \\
            & LRG2 & 0.9969 & $-0.0031$ & 0.0272 & 0.0250 \\
            & LRG3 & 0.9919 & $-0.0081$ & 0.0275 & 0.0361 \\
            & ELG2 & 1.0340 & $+0.0340$ & 0.0422 & 0.0427 \\
            & QSO  & 1.0776 & $+0.0776$ & 0.0374 & 0.0551 \\
        \addlinespace
        2BB & LRG1 & 1.0206 & $+0.0206$ & 0.0426 & 0.0434 \\
            & LRG2 & 1.0043 & $+0.0043$ & 0.0305 & 0.0230 \\
            & LRG3 & 1.0133 & $+0.0133$ & 0.0328 & 0.0395 \\
            & ELG2 & 1.0414 & $+0.0414$ & 0.0480 & 0.0520 \\
            & QSO  & 1.0401 & $+0.0401$ & 0.0337 & 0.0528 \\
        \midrule
        \multicolumn{6}{l}{\bf POST} \\
        \midrule
        1BB & LRG1 & 1.0200 & $+0.0200$ & 0.0325 & 0.0404 \\
            & LRG2 & 0.9713 & $-0.0287$ & 0.0267 & 0.0301 \\
            & LRG3 & 0.9684 & $-0.0316$ & 0.0265 & 0.0282 \\
            & ELG2 & 1.0204 & $+0.0204$ & 0.0383 & 0.0478 \\
            & QSO  & 1.0744 & $+0.0744$ & 0.0395 & 0.0548 \\
        \addlinespace
        2BB & LRG1 & 1.0215 & $+0.0215$ & 0.0337 & 0.0403 \\
            & LRG2 & 0.9842 & $-0.0158$ & 0.0303 & 0.0260 \\
            & LRG3 & 0.9909 & $-0.0091$ & 0.0294 & 0.0307 \\
            & ELG2 & 1.0214 & $+0.0214$ & 0.0367 & 0.0553 \\
            & QSO  & 1.0463 & $+0.0463$ & 0.0369 & 0.0562 \\
        \bottomrule
    \end{tabular}
    \caption{Recovery of $\alpha$ from 25 \texttt{Abacus} mocks for the 1BB and 2BB broadband models, before and after reconstruction. The columns give the posterior means $\langle\alpha\rangle$, its offset from the expected value of unity, the mean posterior uncertainty $\langle\sigma_\alpha\rangle$ and $\sigma(\alpha)$ is the sample standard deviation of $\alpha$ across the mocks. Here, we do not remove any flagged mocks as in Table \ref{tab:mock_alpha_summary}.}
    \label{tab:1bb2bbtable}
\end{table}

\subsection{Recovery of $\alpha$ values}\label{ssec:recoveryofalpha}
\begin{figure}
    \centering
    \includegraphics[width=\linewidth]{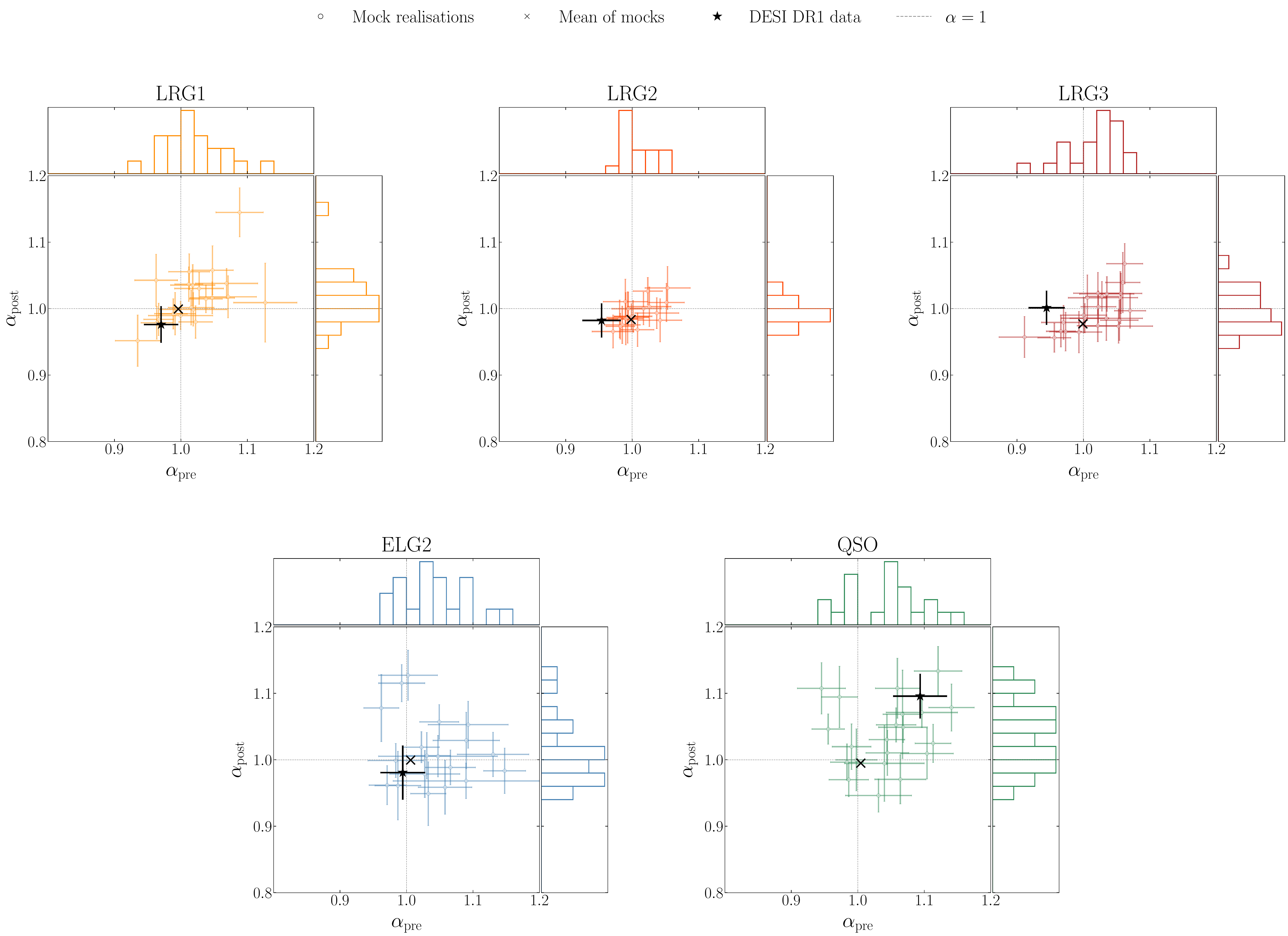}
    \caption{The distribution of recovered $\alpha_{{\rm pre}}$ and $\alpha_{{\rm post}}$ for each tracer.  Open points in the main panels correspond to one mock realisation, and the histograms show the one-dimensional distributions of $\alpha_{{\rm pre}}$ (top) and $\alpha_{{\rm post}}$ (right). The cross shows the result of fitting the mean $w(\theta)$ of the 25 \abacus mocks. The star shows the corresponding DESI DR1 measurement for that tracer. We also plot $\alpha_{{\rm pre}} = 1 $ and $\alpha_{{\rm post}} = 1$ in grey.}
    \label{fig:alphamocks}
\end{figure}

In this section, we test how well the fitting pipeline recovers the transverse dilation parameter on the \abacus mocks, and what spread and uncertainty we should expect for a single DESI-like realisation. We write $\alpha_{{\rm pre}}$ and $\alpha_{{\rm post}}$ for fits to the measurements before and after reconstruction.

Figure \ref{fig:alphamocks} shows the recovered $\alpha_{{\rm pre}}$ and $\alpha_{{\rm post}}$ for each tracer. We apply exactly the same fitting procedure to the mocks and to the data, using the joint fit of Section \ref{ssec:baofitting} in which a single $\alpha$ is shared across all the redshift bins of a tracer sample (Equation \ref{eq:theta_joint}). Each open point in the main panels corresponds to one mock realisation, and the histograms show the one-dimensional distributions of $\alpha_{{\rm pre}}$ (top) and $\alpha_{{\rm post}}$ (right). The cross shows the result of fitting the mean $w(\theta)$ of the 25 \abacus mocks, which should not be confused with the mean of the individual fits. The star shows the DESI DR1 measurement. 

To summarise the individual mock fits, we compute two quantities for each mock. The first is the posterior mean $\alpha_{{\rm mean}}$. The second is $\sigma_{\alpha,68}$, defined as half the width of the 68 $\%$ equal-tail posterior interval. Table \ref{tab:mock_alpha_summary} reports the average of each quantity over the retained mocks, $\langle\alpha_{{\rm mean}}\rangle$ and $\langle\sigma_{\alpha,68}\rangle$, together with the sample standard deviation of the posterior means across the retained mocks, $\sigma(\alpha_{{\rm mean}})$. The quantity $\langle\sigma_{\alpha,68}\rangle$ is the typical uncertainty that a single realisation assigns to itself, while $\sigma(\alpha_{{\rm mean}})$ measures how much the recovered values actually vary from one realisation to another. Only mocks classified as a detection, following the criterion described in Appendix \ref{app:detectionsigmocks}, are included, and any mock flagged in either the pre- or post-reconstruction case is removed from both.

Fitting the mean of the mocks and fitting the individual realisations serve two complementary purposes. Averaging over 25 realisations strongly suppresses the mock-to-mock fluctuations and gives a much cleaner BAO feature. The fit to the mean therefore tests whether the model itself is able to recover $\alpha$ in an unbiased way, since the fiducial cosmology underlying the mocks is the same as that used to compute the template.  We recover this for LRG1, ELG2 and QSO both before and after reconstruction, and for LRG2 and LRG3 before reconstruction. After reconstruction, the fits to the mean of the LRG2 and LRG3 mocks are shifted to slightly lower values, by approximately 0.015 and 0.023, respectively. These shifts are smaller than $\sigma(\alpha_{{\rm mean}})$ for the same samples (0.018 and 0.029), so any residual model bias of this size has a subdominant effect on a single measurement such as DR1. The individual realisations quantify the scatter expected in a single DESI-like measurement. Reconstruction moves the mean of the posterior means closer to unity for every tracer. From Table \ref{tab:mock_alpha_summary}, the offset $\langle\alpha_{\rm mean}\rangle - 1$ changes from $+0.020$ to $+0.018$ for LRG1, from $+0.008$ to $-0.007$ for LRG2, from $+0.016$ to $-0.004$ for LRG3, from $+0.041$ to $+0.013$ for ELG2 and from $+0.041$ to $+0.037$ for QSO, post-reconstruction. 

\begin{figure}
    \centering
    \includegraphics[width=\linewidth]{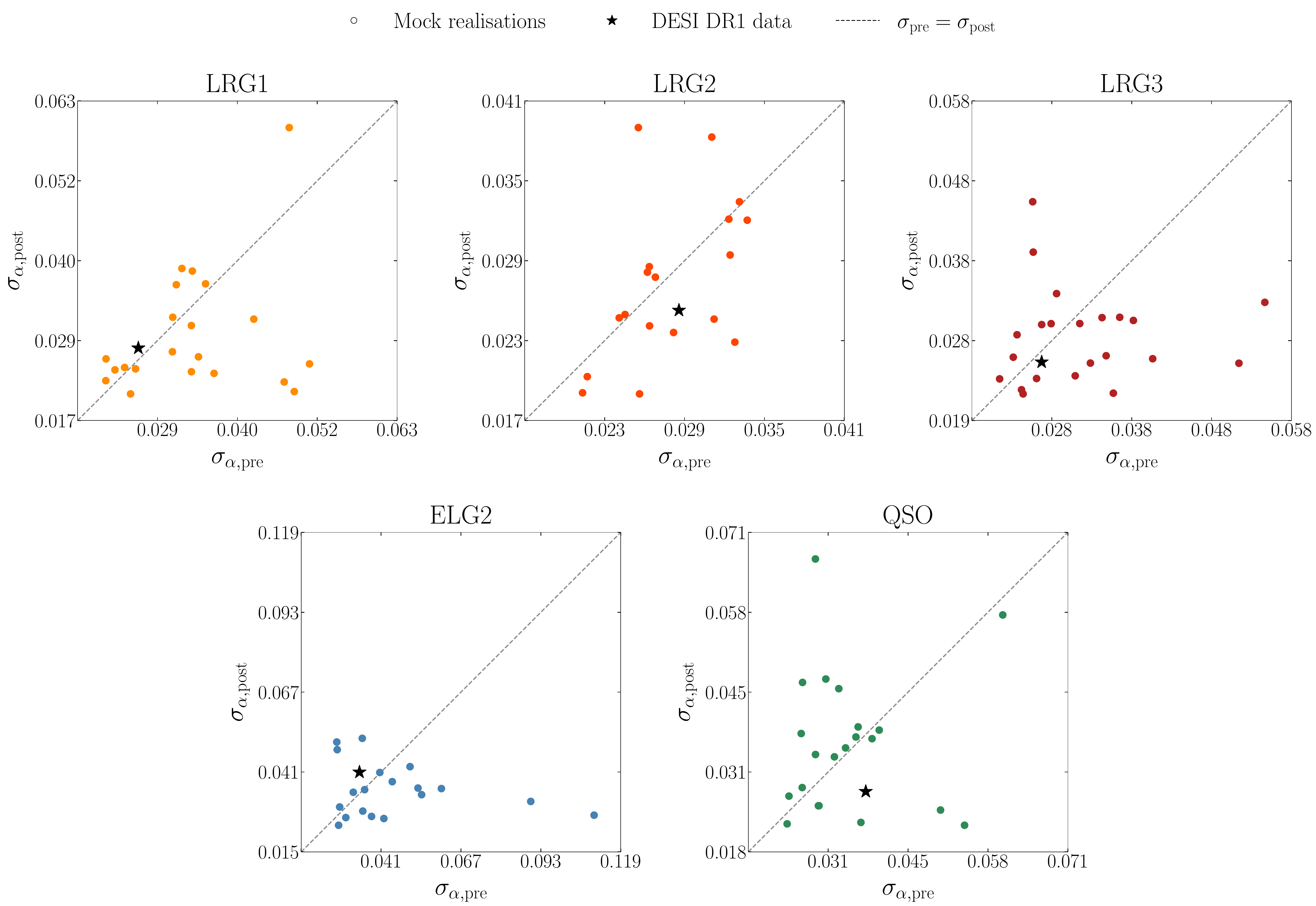}
    \caption{The distribution of errors  on $\alpha_{{\rm pre}}$ and $\alpha_{{\rm post}}$ for each tracer. The star shows the corresponding DESI DR1 measurement for that tracer. We also plot the $\sigma_{\alpha_{{\rm pre}}} = \sigma_{\alpha_{{\rm post}}}$ in grey. }
    \label{fig:sigmaalphamocks}
\end{figure}

Figure \ref{fig:sigmaalphamocks} compares the posterior uncertainty on $\alpha$ before and after reconstruction for each retained mock, together with the DESI DR1 measurement. For every tracer, the mock realisations lie on both sides of the one-to-one line, so reconstruction does not reduce the uncertainty in every realisation. The largest improvements occur for the realisations with the largest pre-reconstruction uncertainty, which lie well below the line. In every panel, the DR1 measurement lies within the region spanned by the mocks, and we conclude that the behaviour of the data is therefore consistent with that expected for a single realisation.

\begin{table}
    \centering
    \begin{tabular}{lccccc}
        \hline
        Tracer & $\langle \alpha_{\rm mean} \rangle$ & $\sigma(\alpha_{\rm mean})$ & $\langle \sigma_{\alpha,68} \rangle$ & $\langle \alpha_{\rm bf} \rangle$ & $\chi^2_{\rm bf}/\mathrm{dof}$ \\
        \hline
        \multicolumn{6}{l}{\textbf{PRE}} \\
        \hline
        LRG1 & 1.0200 & 0.0458 & 0.0327 & 1.0076 & 125.44 / 121 \\
        LRG2 & 1.0081 & 0.0255 & 0.0278 & 1.0036 & 127.04 / 113 \\
        LRG3 & 1.0160 & 0.0413 & 0.0315 & 0.9983 & 151.81 / 144 \\
        ELG2 & 1.0412 & 0.0529 & 0.0451 & 1.0127 & 203.49 / 169 \\
        QSO & 1.0409 & 0.0560 & 0.0336 & 1.0123 & 412.84 / 432 \\
        
        \hline
        \multicolumn{6}{l}{\textbf{POST}} \\
        \hline
        LRG1 & 1.0175 & 0.0402 & 0.0289 & 1.0102 & 126.56 / 121 \\
        LRG2 & 0.9928 & 0.0183 & 0.0272 & 0.9885 & 130.81 / 113 \\
        LRG3 & 0.9957 & 0.0294 & 0.0281 & 0.9835 & 159.60 / 144 \\
        ELG2 & 1.0130 & 0.0520 & 0.0328 & 1.0002 & 182.54 / 169 \\
        QSO & 1.0365 & 0.0507 & 0.0361 & 0.9911 & 429.60 / 432 \\
        \hline
    \end{tabular}
    \caption{Summary of mock $\alpha$ constraints after removing flagged mocks. $\langle \alpha_{\rm mean} \rangle$ is the mean of the posterior means across retained mocks, $\sigma(\alpha_{\rm mean})$ is the standard deviation of those posterior means across retained mocks, and $\sigma_{\alpha,68}$ is defined for each mock as half the width of the 68\% equal-tail posterior interval. $\langle \alpha_{\rm bf} \rangle$ and $\chi^2_{\rm bf}$ are averaged over the minimizer outputs across retained mocks. For each tracer, any mock flagged in either PRE or POST is removed from both.}
    \label{tab:mock_alpha_summary}
\end{table}

\subsection{Detection significance on mocks}\label{app:detectionsigmocks}

\begin{figure*}
    \centering
    \includegraphics[width=0.30\textwidth]{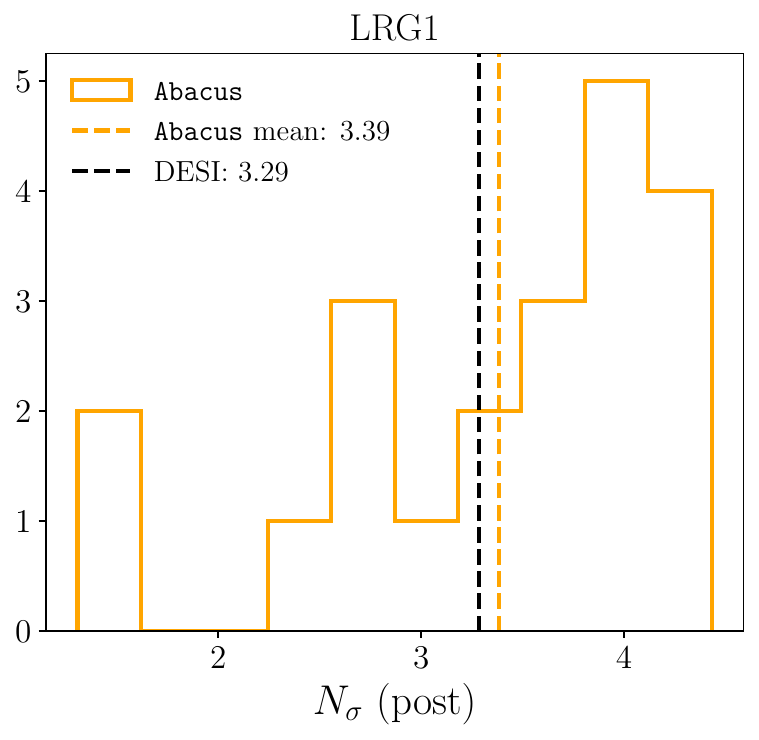}
    \hfill
    \includegraphics[width=0.30\textwidth]{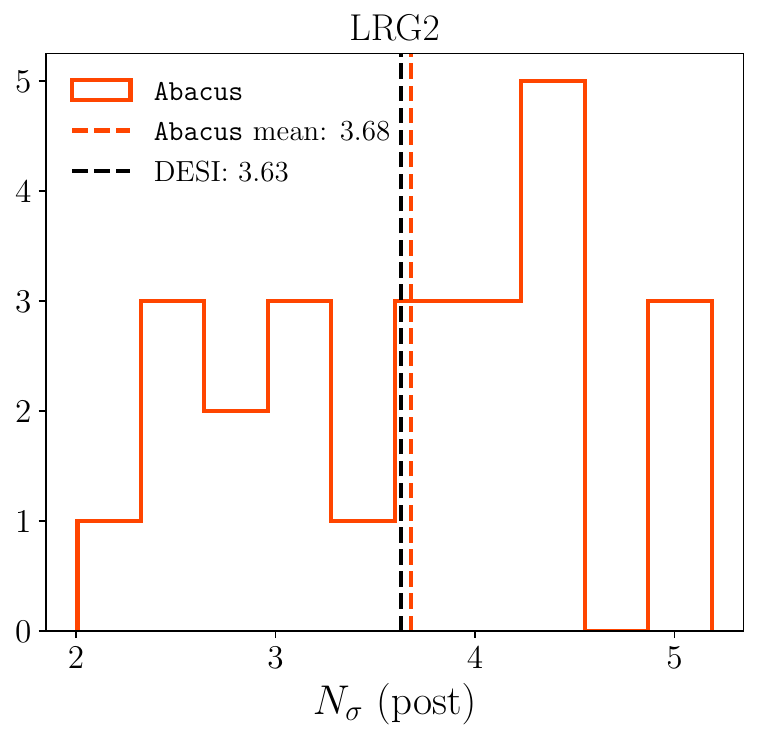}
    \hfill
    \includegraphics[width=0.30\textwidth]{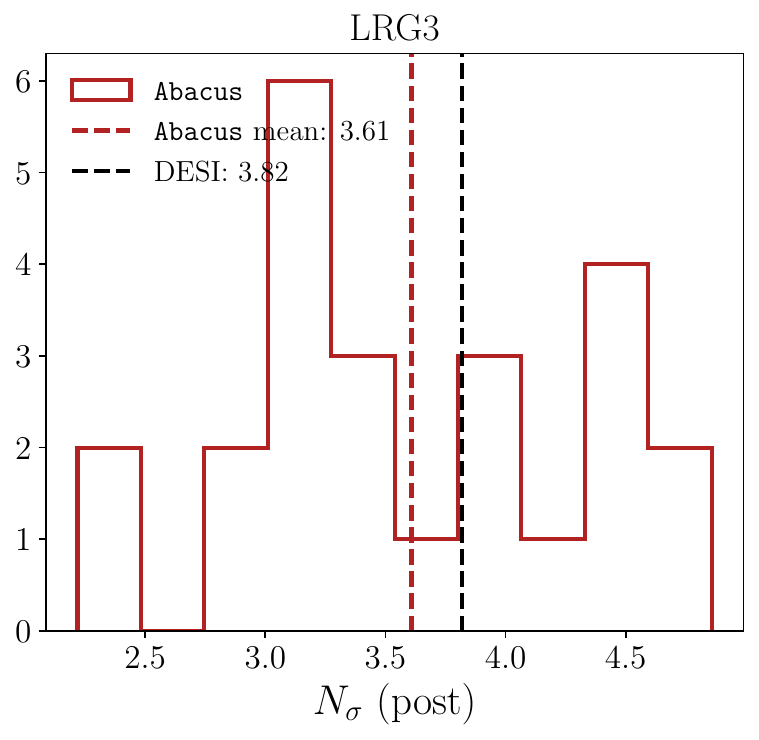}
    \vspace{0.5cm}
    \makebox[\textwidth][c]{%
        \includegraphics[width=0.30\textwidth]{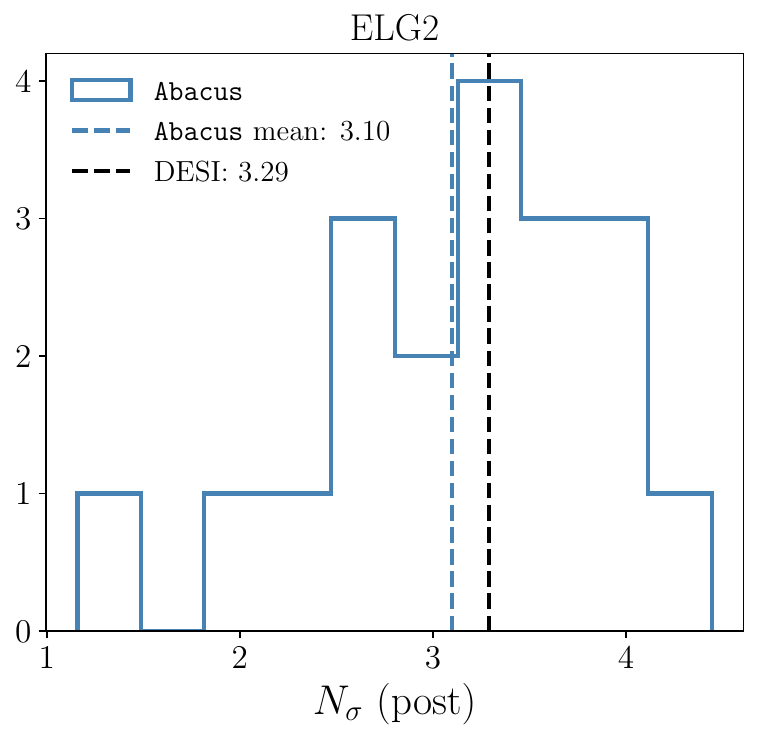}
        \hspace{0.03\textwidth}
        \includegraphics[width=0.30\textwidth]{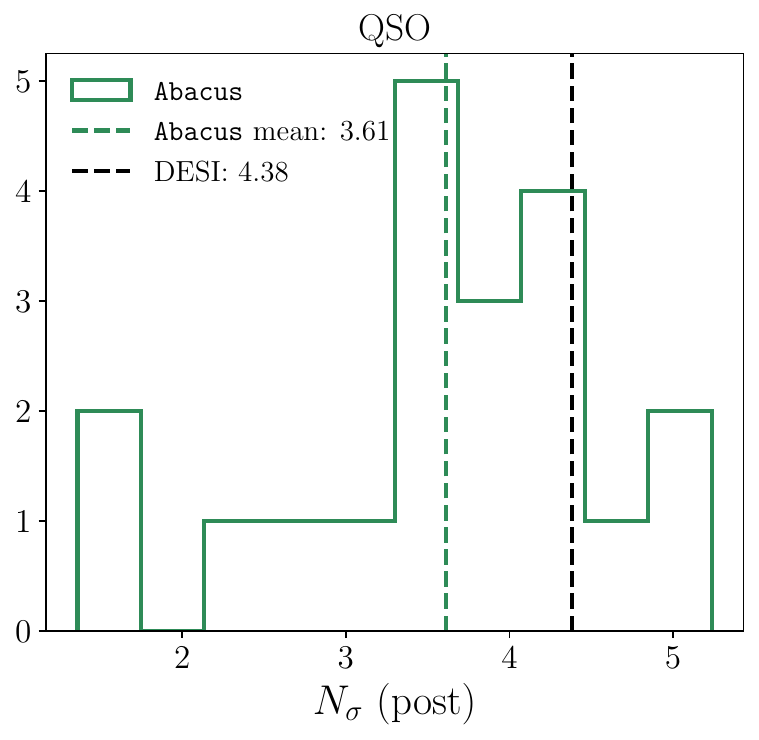}
    }
    \caption{Comparison of $N_\sigma$ between the data and mock catalogues for LRG1, LRG2, LRG3, ELG2 and QSO (post-reconstruction).}
    \label{fig:nsigma}
\end{figure*} 

In this section, we quantify the expected detection significance that we obtain from the mocks.\footnote{We refer the reader to Section \ref{ssec:detectionsig} for how the detection is quantified and how the $\chi^2$ statistic is used.} However, not every mock realisation yields a usable measurement of $\alpha$. When the BAO feature in a single realisation is weak, the posterior on $\alpha$ can run into the edge of the flat prior $\alpha \in [0.8, 1.2]$. In that case, the posterior mean and width are set largely by the prior rather than by the data, and including such mocks would distort the statistics in Table \ref{tab:mock_alpha_summary}. We therefore classify a mock as a detection only if it satisfies two conditions:

\begin{enumerate}
    \item The interval $\alpha \pm \sigma_\alpha$ lies entirely within the prior.
    \item $\Delta\chi^2 \geq 0$, so that the BAO model fits the realisation at least as well as the no-BAO model. We define $\Delta\chi^2$ in Equation \ref{eq:deltachi2}.
\end{enumerate}  
The first condition removes realisations whose posterior is truncated by the prior, and the second removes realisations in which the no-BAO feature is preferred. A mock that fails any condition in either the pre- or post-reconstruction fit is removed from both, so that the pre- and post-reconstruction statistics are computed from the same set of realisations. In the post-reconstruction case, this removes 2 mocks for LRG1, 5 for LRG2, 2 for LRG3, 3 for ELG2 and 4 for QSO. In the pre-reconstruction case, this removes 3 mocks for LRG1, 4 for LRG2, 2 for LRG3, 3 for ELG2 and none for QSO.

Figure \ref{fig:nsigma} shows the distribution of $N_\sigma$ after reconstruction across the retained \abacus mocks, together with the value obtained for DESI DR1 in this work. The mean significance over the mocks is 3.39, 3.68 and 3.61 for LRG1, LRG2 and LRG3, and 3.10 and 3.61 for ELG2 and QSO. The corresponding DR1 values are 3.29, 3.63, 3.82, 3.29 and 4.38. For the LRG samples and ELG2 the DR1 significance differs from the mock mean by at most 0.21, which is small compared with the spread of $N_\sigma$ across realisations. The largest difference is for QSO, where DR1 gives a detection 0.77 above the mock mean. This value still lies within the range spanned by the individual QSO mocks, and is consistent with an upward fluctuation of the BAO feature in a single realisation of a sparse, noise-dominated sample. The detection significance in the data is therefore in agreement with that of the individual mock realisations for every tracer. Together with the agreement in $\alpha$ and $\sigma_\alpha$ shown in Figures \ref{fig:alphamocks} and \ref{fig:sigmaalphamocks}, this indicates that the DR1 measurement behaves as a typical realisation, and that the performance of the pipeline established on the mocks applies to the data.

\section{Fits to individual redshift bins}\label{app:individualfit}

As a consistency check to the joint fit, we also fit each tomographic bin separately, keeping the same template, broadband model, fitting range and priors as in the fiducial analysis but allowing a separate $\alpha$ in every bin, as specified in Section \ref{ssec:baofitting}. 

\begin{figure}
    \centering
    \includegraphics[width=\linewidth]{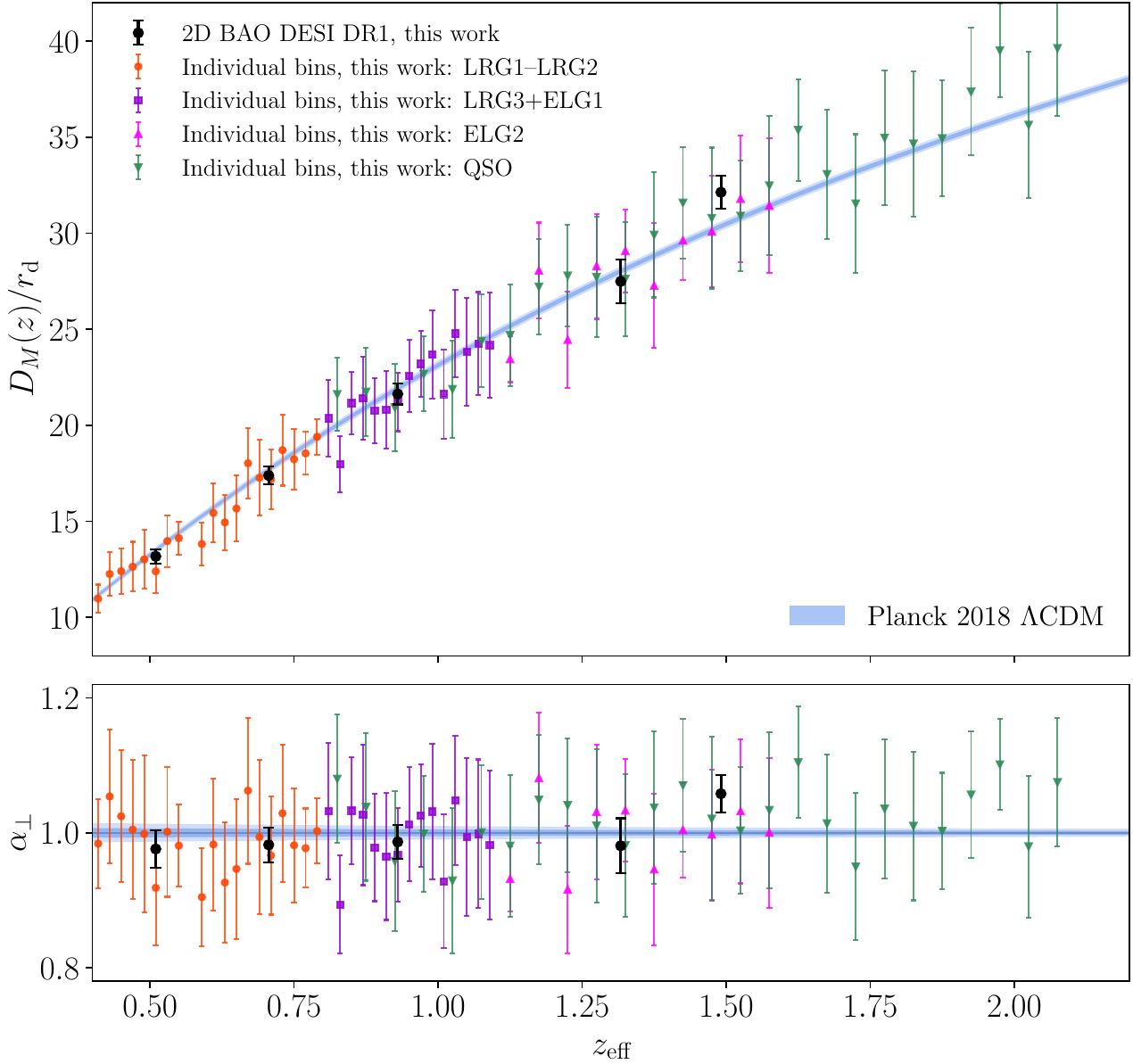}
    \caption{{\it Top panel}: We plot $D_M(z)/r_{\rm d}$ from the individual bin fits of all tracers, together with the joint measurements of this work on DESI DR1 data and the Planck 2018 $\Lambda$CDM prediction. {\it Bottom panel}: The same fits as the top panel but in terms of $\alpha$. The dark orange points indicate the individual fits for the LRG1 and LRG2 sample ($0.4 < z < 0.8$), the dark violet square points indicate the fits for the LRG3$+$ELG1 sample ($0.8 < z < 1.1$), the magenta upward triangle indicates the fits for the ELG2 sample ($1.1 < z < 1.6$) and the downward green triangle indicates the fits for the QSO sample ($0.8 < z < 2.1$).}
    \label{fig:individualbins}
\end{figure}

The top panel of Figure~\ref{fig:individualbins} shows $D_M(z)/r_{\rm d}$ from the individual bin fits of all tracers, together with the joint measurements of this work and the Planck 2018 $\Lambda$CDM prediction. Taken together, the individual bins trace the redshift evolution of $D_M(z)/r_{\rm d}$ across the full range $0.4 < z < 2.1$ and follow the shape of the fiducial prediction. They scatter around the joint measurement of their own tracer, and bins from different tracers agree where they overlap in redshift, as for LRG and QSO at $0.8 < z < 1.1$ and for ELG2 and QSO at $1.1 < z < 1.6$. The uncertainties on the individual bins are substantially larger than those of the joint measurements, as expected, since each bin contains only a fraction of the galaxies of its tracer sample and the BAO feature in a single thin shell is measured at lower signal-to-noise. The joint fit combines the information from all the bins of a tracer through a single $\alpha$, which gives the correspondingly smaller uncertainty.

The bottom panel shows the same fits in terms of $\alpha$, which isolates the deviation of each bin from the fiducial cosmology. This is most relevant for the QSO sample, which spans the widest redshift range, $0.8 < z < 2.1$, and therefore a natural concern is whether a single $\alpha$ to extract the BAO is robust. Fitting it with a single $\alpha$ assumes that the ratio of the true to the fiducial $D_M(z)/r_{\rm d}$ does not change appreciably across the sample. The Hubble constant and the sound horizon only rescale this ratio by a redshift-independent factor, so any evolution with redshift arises only from a difference in the shape of $D_M(z)$ between the true and fiducial cosmologies. The individual bin fits test this directly, since each bin has its own $\alpha$ and therefore is a local measurement. If a single $\alpha$ were not adequate, the individual QSO values of $\alpha$ would drift systematically as the redshift increases. Instead, they scatter around the joint QSO value across the full redshift range with no visible trend. We therefore conclude that a single $\alpha$ is an adequate description of the QSO sample at the precision of DR1.

\section{Off-block covariance and parameter inference}\label{app:offblock}

We decompose the QSO mock covariance as $\Cmat = {\sf D} + {\sf N}$, where ${\sf D}$ contains the diagonal blocks associated with each redshift shell and ${\sf N} = \Cmat - {\sf D}$ contains the elements connecting distinct shells. All quantities in this appendix use the raw mock covariance from the 1000 EZ FFA mocks, before any Hartlap correction, to isolate the effect of ${\sf N}$.

\subsection{Compatibility of the off-block covariance with sampling noise}

If the true covariance is block diagonal, the off-block elements of the sample covariance arise purely from finite-mock sampling fluctuations. Under this null hypothesis, and assuming independent Gaussian mocks, the whitened off-block matrix
\begin{equation}
    W = {\sf D}^{-1/2}\, {\sf N}\, {\sf D}^{-1/2}
\end{equation}
has uncorrelated elements with rms $1/\sqrt{N_{\rm mocks}-1}$. We compare the norm of the observed ${\sf W}$ with its distribution over 200 covariance realisations drawn from a Wishart distribution with ${\sf D}$ as the null covariance and $N_{\rm mocks}-1 = 999$ degrees of freedom. The observed value is 15.89, against $15.85 \pm 0.04$ under the null, with a probability-to-exceed of $p = 0.14$. The off-block covariance is therefore compatible with finite-mock sampling noise.

\subsection{Projection onto parameter space}

The exact change in the precision matrix associated with ${\sf N}$ is
\begin{equation}
    \Omega_{\sf N} \equiv {\sf D}^{-1} -{\sf C}^{-1},
\end{equation}
which requires no assumption that ${\sf N}$ is small. Although ${\sf N}$ is individually small, $\Omega_{\sf N}$ need not be, since matrix inversion propagates the effect of ${\sf N}$ throughout the precision matrix. Within the diagonal blocks, we find a median fractional difference of $-0.62$ between ${\sf D}^{-1}$ and ${\sf C}^{-1}$.

The likelihood, however, depends on the precision matrix only through its contraction with the model. Defining the Jacobian $J_{ia} = \partial t_i / \partial \theta_a$ of the model ${\bf t}$ with respect to the parameters $\boldsymbol\Theta$, the Fisher matrices for the two covariances are
\begin{equation}
    {\sf F}_{\sf C} = {\sf J}^{\!\top} {\sf C}^{-1} {\sf J}, \qquad {\sf F}_{\sf D} = {\sf J}^{\!\top} {\sf D}^{-1} {\sf J},
\end{equation}
so that
\begin{equation}
    {\sf F}_D - {\sf F}_C = {\sf J}^T \Omega_N {\sf J}.
\end{equation}
The columns of ${\sf J}$ span the directions in data space along which the model can move. For the QSO sample, with 511 data points and 79 parameters, $\Omega_{\sf N}$ can differ substantially from zero in directions that the parameters cannot reach, without affecting the inference. Its impact on the parameter uncertainties is controlled by ${\sf J}^T\Omega_{\sf N} {\sf J}$, and its impact on the central values by ${\sf J}^T \Omega_N {\bf u}$, where ${\bf u}$ is the data residual.

We evaluate ${\sf J}$ numerically at the best fit. The marginalised Fisher uncertainty on $\alpha$ is $\sigma_\alpha = 0.0225$ for ${\sf C}$ and $0.0307$ for ${\sf D}$, so that ${\sf N}$ increases the apparent precision on $\alpha_\perp$. The corresponding shift of the best-fit parameters when ${\sf N}$ is included is, for a locally linear model,
\begin{equation}
    \delta\boldsymbol\theta_N = -{\sf F}_C^{-1} {\sf J}^T \Omega_N {\bf u}_D,
\end{equation}
where ${\bf u}_D$ is the residual at the best fit obtained with ${\sf D}$. Its $\alpha$ component is $-0.0057$, or $0.25\sigma$. Over all parameters jointly, the shift is dominated by the amplitude and broadband nuisance parameters.

The off-block covariance therefore barely moves the recovered BAO scale, but increases the apparent precision on it. Applying the appropriate correction to each covariance, $\psi = 0.49$ for ${\sf C}$ and $\psi_i \simeq 0.98$ per block for ${\sf D}$, gives uncertainties of $0.032$ and $0.031$ respectively, in agreement. Both prescriptions also recover the input cosmology on the mean of the \abacus mocks, with $\alpha = 0.995$ and $0.997$. The adopted treatment, the full mock covariance with the standard Hartlap correction, is therefore robust, and the large correction for the QSO sample reflects the removal of spurious information rather than a loss of genuine constraining power.

\bibliographystyle{JHEP}
\bibliography{biblio.bib}

\end{document}

%% file: abstract.tex
We present the angular BAO analysis using DESI DR1 data as a complementary approach to DESI's primary 3D BAO analysis. We perform the analysis using the two-point angular correlation function $w(\theta)$ in tomographic redshift shells of $\delta z = 0.02, 0.05$, spanning the redshift range $0.4 < z < 2.1$ using the DESI tracers: LRG1, LRG2, LRG3+ELG1, ELG2 and QSO. The main benefit is that it requires no fiducial cosmology to convert redshifts into distances and yields a direct measurement of the transverse dilation parameter $\alpha_{\perp}(z)$. We fit each tracer with a single shared $\alpha_{\perp}(z)$ using templates constructed in a fiducial cosmology and a mock-based covariance from 1000 EZmock realisations. We measure $\alpha_\perp$ to $2.5$ -- $4.2\%$ precision in each tracer bin. Compared to the $D_M/r_{\rm d}$ measurements obtained from the DESI DR1 3D analysis for the same tracers, our results are consistent within the statistical precision. We find no evidence for a shift in the transverse BAO scale between the 2D and 3D analyses. Flat $\Lambda$CDM fits to $\alpha_\perp(z)$ alone give $\Omega_{\rm m} = 0.198^{+0.045}_{-0.068}$, $1.6\sigma$ below the DESI 3D value, and the pull is dominated by the $z =1.49$ QSO bin, whose $\alpha_\perp(z=1.49)=1.058\pm0.028$ sits $\sim 2\sigma$ above the fiducial. Restricting to LRGs and ELGs only, our $\Omega_{\rm m}$ and $H_0 r_{\rm d}$ agree with the 3D transverse-only fit over the same bins to better than $0.1\sigma$. With an additional BBN prior, the full 2D sample gives $H_0 = 63.9^{+1.9}_{-3.3}\, {\rm km\,s^{-1}\,Mpc^{-1}}$. This pipeline is implementable on forthcoming DESI data releases, where the increase in volume and sampling density will determine if the QSO offset persists and will also tighten 2D BAO constraints.